\documentclass[preprint,aps,amsmath,nofootinbib]{revtex4}
\usepackage{graphics,color,array,dcolumn}
\usepackage{graphicx}
\usepackage{calc}

\usepackage{amsmath}
\usepackage{float}
\usepackage{amssymb}
\usepackage{xspace}
\allowdisplaybreaks[1]

\DeclareMathOperator{\acos}{cos^{-1}}
\newlength{\figurewidth}
\newcommand{\beq}{\begin{equation}}
\newcommand{\eeq}{\end{equation}}
\newcommand{\bea}{\begin{eqnarray}}
\newcommand{\eea}{\end{eqnarray}}
\newcommand{\ba}{\begin{array}}
\newcommand{\ea}{\end{array}}

\newcommand{\mn}{{\mu\nu}}
\newcommand{\mnp}{{\mu\nu^\prime}}
\newcommand{\pt}{\partial}
\newcommand{\pd}{(2\pi)^d}
\newcommand{\al}{\alpha}
\newcommand{\bt}{\beta}
\newcommand{\g}{\gamma}

\newcommand{\ta}{\theta}
\newcommand{\lam}{\lambda}
\newcommand{\LAM}{\Lambda}
\newcommand{\G}{\Gamma}
\newcommand{\nb}{\nabla}
\newcommand{\de}{\delta}
\newcommand{\D}{\Delta}
\newcommand{\OM}{\Omega}

\newcommand{\sg}{\sigma}

\makeatother
\begin{document}
%
\title{Green's function of the Vector fields on DeSitter Background}
\setlength{\figurewidth}{\columnwidth}
%
\author{Gaurav Narain}
\email{gauravn@nu.ac.th}
\affiliation{
The Institute for Fundamental Study ``The Tah Poe Academia Institute",
\\ Naresuan University, Phitsanulok 65000, Thailand.}
%
%
\begin{abstract}
In this paper we study the propagator of a vector fields on a 
euclidean maximally-symmetric background in arbitrary space-time dimensions. 
We study two cases of interest: 
Massive and massless vector fields. In each case we 
computed the propagator of the vector fields on euclidean deSitter background, 
isolating the transverse and longitudinal part. 
In both case of massive and massless vector fields, 
the short distance limit of the full propagator agrees 
with the flat space-time propagator.
In the case of massive propagator, the transverse part has a well 
defined massless limit, and in this limit it goes to the 
transverse propagator for the massless fields.
The transverse propagator for antipodal point separation 
is nonzero and negative, but vanishes in the flat 
space-time limit (Ricci scalar going to zero). 
The longitudinal part of the massive propagator diverges 
as $1/m^2$, where $m$ is the mass of the field.
The longitudinal part of the massless propagator is gauge 
dependent and in particular is proportional to the gauge 
parameter used in the gauge fixing condition. It vanishes 
in the Landau gauge. Comparison with the past literature 
is made. 
\end{abstract}

\maketitle
%
%

\section{Introduction}
\label{intro}

The Big-bang model of cosmological evolution although explains the
expansion of the universe to a great accuracy, but the 
observations also raises puzzling issues, which are better 
known in cosmology as Horizon and flatness problem. 
A possible solution to these problems is to have 
an era of exponential expansion in the early stages of the 
universe, which is commonly known as Inflation 
\cite{Guth1980, Linde1981, Starobinsky1982}. 
It turns out that such an era of exponential expansion 
can also offer a possible explanation for the large scale
structure of the universe. A possible theory is that the large scale 
structures perhaps might have got their origin from the 
quantum fluctuations in the early stages of universe, which got 
amplified during the exponential expansion. But, at the moment there is no
detailed particle physics mechanism known which is responsible for
inflation. However there are several models, which try to get an era of 
exponential expansion of the space-time, which should last at least 
sixty $e$-folds. One possible way of getting an era of exponential 
expansion theoretically is to assume that the universe is filled by a perfect fluid, 
having a constant energy density throughout the space-time. This could be 
the vacuum energy of the space-time, also known as cosmological constant
denoted by $\LAM$. This kind of fluid has equation of state $p=-\rho$, thereby
giving negative pressure. Einstein's equation then tells that this scenario
will result in accelerated expansion of the space-time.
If the universe is assumed to be maximally symmetric, then this will
imply an exponential growth of the scale factor for the conformally 
flat space-time. This is a deSitter space-time. A phase of accelerated 
expansion in the early universe also explains why we notice almost 
equal temperatures in the CMB even for points which are very far apart, 
with very small amount of anisotropies. 

A DeSitter type of expansion was not only witnessed during the 
inflationary era, but it has been noticed from the recent cosmological 
observations that the universe at present is witnessing an 
accelerated expansion. It was first noticed in the observations of 
type 1a supernova \cite{Riess}. More observations conducted later 
independently by other groups further confirmed the 
inference of accelerated expansion of the universe. 
CMB anisotropies from the WMAP data, tells that the universe is 
close to flat, and to account for this flatness one needs around
$70\%$ of dark energy. Planck's data further confirmed this 
thereby giving more accurate estimate on the percentage of dark energy. 
Independent observations of
large scale structures further established the knowledge that 
universe is accelerating and consist of roughly $70\%$ of dark energy. 
One simplest and possible explanation for these is the 
acceleration being driven by vacuum energy (same as for inflation). 
This vacuum energy has an equation of
state $p=-\rho$, thereby giving negative pressure resulting in accelerated 
expansion of the universe. To date this is the best explanation for the
dark energy, fitting a variety of other observations to a great accuracy as well. 
However, there have been other proposals also for explaining this: 
for example Quintessence etc. In either case, one is trying to 
explain the late time acceleration of the universe by building a 
sensible theory for dark energy. In these times of accelerated expansion, 
it is expected that the universe will again have deSitter kind of geometry.  

Although the observations tells that the early and late time universe 
has witnessed and is witnessing an era of accelerated expansion
respectively, which can be modelled to a good approximation with 
a deSitter kind of geometry, but we still do not have a good 
physical theory to explain any of them. While the former
is explained using inflationary scenarios, and the later using dark energy models,
however in either case we are still lacking a good understanding as to what may
be the cause of either of them? We do not know why they are happening, 
how they are happening and why they started?
In either case we know from observations that they have either happened or is 
happening, and we have some models which try to explain either of them
but a complete physical picture is still missing. This 
demands for further study. Mathematically consistent quantum theories 
explaining either of them is the required goal, but till now 
our efforts have yet not been rewarding enough. 

In the last century, we witnessed two important major revolution in physics:
general relativity and quantum mechanics. Although we have been largely 
successful in constructing a theory describing relativistic quantum particles,
called quantum field theory, which has been very accurate in explaining 
the observations from high-energy scattering and condensed matter experiments,
but so far our efforts to combine gravity in this picture has been largely in vain.
While this has led to an extensive research program on quantum gravity: 
a theory which successfully combines quantum mechanics and general relativity, 
but till date we are yet to come across a sensible theory which 
doesn't have any problems. However, there are many approaches 
to Quantum gravity which try to put successfully quantum mechanics 
and gravity together, but in all these approaches there are some 
shortcomings, forcing the researchers to be skeptic to either of them. 
In these situations of turmoil, it is good to study quantum matter fields
on a curved background first, to get an understanding of how background
curvature of space-time might be affecting the quantum phenomenas 
between the matter fields, while the energy is still not large enough to 
demand quantisation of gravity. This has lead to an extensive amount 
of research in the direction of quantum field theory on 
curved background \cite{Birrell1982}.
While this may not be mathematically consistent fully \cite{Duff1980},
but it may still give some physical insight in to physics happening 
at those energy scales. Moreover such studies will also tell 
about the energy scales up to which this particular theory 
can be trusted, and when it starts to break down 
thereby giving unexpected results?

With this in mind we study the simplest of all the background, 
conformally flat deSitter space-time. This is a maximally symmetric
space-time and next in simplicity to flat space-time. 
While this is also physical, due to reasons mentioned above, it 
becomes a good theoretical laboratory to construct and understand 
methods of quantum field theory on curved backgrounds. 
Due to this reason, a large amount of effort has gone in to
studying quantum matter fields on deSitter space-time. 
As in flat space-time the first and important step in constructing
a successful QFT is to find the vacuum of the theory and 
construct fock states. In curved space time similar direction
was taken, and for simplicity scalar field was considered first
\cite{Chernikov, Tagirov, Candelas1975, Spindel2, Bunch1978, Mottola1984}.
The issue of massive scalar field was further investigated 
in \cite{Burges1984, Allen1985, Allen1987}. It was found that for massive 
scalar there is one-complex parameter family of 
deSitter invariant vacuum states, of which the ``euclidean'' vacuum 
has a special place as for this the propagator has a Hadamard form
for short distances, while for the massless scalar there is no
deSitter invariant vacuum state, and deSitter invariance is
broken, as the propagator exhibits a growing behaviour 
in the infrared. Also it was noticed that the massless limit for the massive
theories is not well defined on a curved background. 
This was also seen earlier in \cite{Ford1977}. 
Having a problematic massless limit of the massive propagator
and a problematic IR divergent massless propagator, 
puts the whole perturbative quantum field theory procedure of 
analysing quantum regime of theories at stake. Therefore it 
becomes a worry, as to how one should be studying processes
involving virtual particles on a deSitter background?
These skepticism regarding QFT on deSitter background,
therefore required attention and effort was made in 
the past to resolve these issues. One scenario that was 
considered in \cite{Folacci1992} was to treat the scalar
theory as a gauge theory, and removes the problematic 
term from the propagator by incorporating a sort of gauge fixing 
term in the action of the theory (this problematic term 
was seen to arise from a zero mode present in the theory).
Then the path-integral of the 
full theory enjoys a BRST symmetry and new propagator is
accordingly computed. This new propagator has a well defined
massless limit and is also regular in the infrared,
in the sense that it doesn't diverge at long-distances. An approach was 
later taken in \cite{Bros2010}, where it was suggested 
that the problematic divergent term of the massive propagator
which diverges in the massless limit, can be subtracted by hand
in order to define a regular propagator which has a well defined 
massless limit. This propagator however exhibits a growing 
behaviour in the IR. Also the massless propagator 
when computed in deSitter exhibits this IR pathology.
This has been a source of concern since then. 
Some possible remedies have been carried out since then. 
It is believed that perhaps this IR problem is artificial and 
is present only at the tree level. In fact it has been shown that if one incorporates a 
small local interaction, then this generates a dynamical mass due to
quantum corrections, thereby resolving the IR problems 
\cite{Burgess2010, Serreau2011, Boyanovsky}. 
This is done either perturbatively or non-perturbatively by summing 
infinite set of cactus diagrams. A non-vanishing mass was 
also witnessed in the stochastic approach in \cite{Starobinsky1994}. 
These computations meant that in the presence of interactions 
it is possible to safely take the massless limit. In \cite{Gautier2013, Youssef2} 
the authors have gone beyond the local approximations and 
considered non-local contributions to self energy of the scalar field,
by summing the Dyson-Schwinger series. Another interesting 
non-local study was conducted in \cite{Hollands2011}. 
In \cite{Youssef2} in an interesting piece of work, it was shown
that in the ladder-rainbow approximation, 
the green's function so obtained doesn't have any IR pathologies.  
Another work which is of relevance is \cite{Marolf2010, Marolf2011}.
While the dust has yet not been settled and demands 
further research even in the case of scalar
\footnote{The problem is important as the massless and IR limit 
of the massive propagator doesn't commute, thereby implying that
even a small amount of curvature, as in the present universe might
lead to drastic puzzling results}, we decided to explore the 
situation in the case of vectors on a deSitter background instead
(see \cite{Akhmedov2013} for nice lecture notes on QFT on 
deSitter background).

Vector fields play a crucial role in the present known picture of the 
standard model of the particle physics. They describe the behaviour
of photon field which is the carrier of electromagnetic force. These
are the simplest vectors in the standard model as their gauge group 
is the smallest. Beside them, vectors are also used in studying the
nature of force carriers in electroweak and strong interactions.
The vectors describing the 
electroweak forces are massive in nature, a mass they 
acquire through Higgs mechanism, while they are massless 
at higher energies. On the other hand the carrier of strong forces
is massless in nature. In either case of electroweak and strong forces
the vectors also carry a gauge index describing their gauge charge (also
called hyper-charge),
and follow a complicated algebra of corresponding Lie group. 
While the present standard model of particle physics is set 
up on a flat background, where it is renormalizable and well 
defined quantum mechanically, its corresponding partner on the 
curved space-time is non-renormalizable and path-integral
describing the quantum theory on curved background is not 
defined. This is matter of worry, as most of the cosmological 
data that we get is in form of photons, which have been living 
in curved space-time. Even though the curvature might be small,
but the fact that it is non-zero makes all the known methods of 
quantum field theory on flat space-time to fall apart. Even though 
we might be far away in attaining a quantum theory of gravity and 
doesn't have any possible way of testing them, but this doesn't 
demotivate us in searching the answers to what might happen to 
the methods of usual flat space-time QFT on a curved background when the 
curvature is not large enough? Given that vectors play a central 
role in the building of standard model, it becomes important 
to know their behaviour during an exponentially expanding era 
in cosmological evolution history. 

For this reason we plan to study vectors on deSitter background.
For simplicity we study first the vectors which doesn't carry 
any gauge index having a nontrivial group structure. In particular
we study two cases of interest: massive and massless vectors
fields on deSitter space-time. This has been studied in past
also. The first study of spin-1 fields on maximally 
symmetric space-time was done in \cite{Katz, Peters, Spindel1}. Subsequently 
it was studied on $CP^n$ in \cite{Warner} and in different gauges 
in \cite{Katz, Peters, Drummond}. A more extensive study on the green's
function on maximally symmetric spaces was conducted in 
\cite{AllenJacob1985}. They considered both massive and massless
vector in euclidean space times and computed the propagator for 
the theory on both deSitter and anti-deSitter background. This was a 
kind of first study and many issues like gauge dependences, infrared 
divergences and massless limits have been left out. While 
in later papers by others these issues have been considered, but 
to date dust remains unsettled on these issues. Recently, due
to the current observational evidences on CMB and dark energy
this direction of research has become more active than before.
Due to which the problem of gauge artefacts in the photon propagator 
have been studied in \cite{Youssef1}. It was found that the 
infrared divergences of the theory are purely gauge artefact and
disappear in the particular gauge choice. The case of massive
vectors have been recently investigated using canonical approach 
in \cite{Higuchi2014}, where it was seen that in the various limits of the Stueckelberg 
parameter (used in their calculations), their results goes to either 
the one given of \cite{AllenJacob1985} or to the one 
given in \cite{Tsamis2006}. It was noticed that
the green's function of the massive vector fields have 
a well defined massless limit and matches with the 
flat space-time expression at short distances.
Massive vectors were also studied 
in \cite{Cotaescu}, where canonical quantisation was used
to obtain the propagator of the theory. These were found to satisfy
similar equations as in \cite{Tsamis2006} but had different 
structure. Massless vector fields were also studied in 
\cite{Garidi2008}, while in \cite{Behroozi2005} the authors 
studied conformally invariant wave equation for massless 
vectors and scalars.

In this paper we compute the propagator of the massive and 
massless vector fields on a deSitter background. While, this 
problem has been studied before, we revisit it here, 
following same methodology of computation both for massive 
and massless vector fields. We use path-integrals to determine the 
equation satisfied by the two-point connected green's function
of the theory. The advantage of this formalism over solving just
the equation of motion is that the methodology can be easily 
extended to take in to account interaction in the quantum field theory
framework in a straight-forward manner. At the tree level, this matches with the equation 
of motion, but under quantum corrections things will be different. 
In this sense the interpretation of the equation satisfied by green's
function changes. Its not just satisfying the equation of motion
but it is more like the inversion of the Hessian of the theory on a deSitter 
background. In this strategy it is more like determining the propagator of 
the virtual particles (somewhat similar strategy was also 
used in \cite{Candelas1975} in his investigation of scalar fields
on deSitter background). We then decompose the
two-point connected green's function into transverse and 
longitudinal part, and set to determine each of them for both the
massive and massless case. The equation satisfied by the transverse 
part of the green's function for both the massive and massless fields 
has the same structure. Furthermore it is noticed that
the transverse massive propagator goes to the transverse 
massless propagator when the mass is taken to zero. 
When the points are antipodal, we get a simple expression 
for the transverse propagator, which might give a wrong indication
that the massless limit will not commute with the antipodal point separation limit.
But this will be misleading, as when one actually takes the massless 
limit, one gets a perfectly well defined answer. 
This is different from the usually witnessed situation in the case 
of scalar fields, where the massless limit is not well defined
\cite{Folacci1992, Bros2010}. 

The longitudinal part has a different story. The longitudinal 
part of the massive case doesn't have a well defined massless limit,
in the sense that it diverges as $1/m^2$. While this is also 
true for flat space-time, where the longitudinal part is seen to 
diverge like $1/m^2$, the short distance limit of the longitudinal 
part of the massive propagator is seen to agree with the longitudinal 
part of the massive flat space-time propagator. The longitudinal
part of the massless propagator is gauge dependent and is 
proportional to gauge parameter. It vanishes in the Landau 
gauge. 

We agree with \cite{Higuchi2014} in some things. In \cite{Higuchi2014}
an extra Stueckelberg parameter has been added, which played actually the
role of gauge-fixing parameter. It was shown there that it is possible to 
take the mass-less limit in a smooth manner in the presence of 
Stueckelberg parameter. But they also mentioned that in this limit 
it is no longer possible for them to take Stueckelberg parameter to 
infinity, as then they witness divergence in the two-point function. 
Although they are successful in taking the massless limit, but 
the fact that they witness a diverging two-point function in the 
limit when Stueckelberg parameter goes to infinity, only indicates 
the problem of divergence has been transferred from one parameter
to the other, which in their case is the Stueckelberg parameter. 
To me this is just the same infrared divergence 
one witnesses in our paper also whose source is actually the presence of zero modes. 
This is actually present in the longitudinal part of the green's function,
as was also noticed in \cite{Higuchi2014}. 

On comparing our results with the past literature, it is noticed that
we agree with the past computations done in 
\cite{AllenJacob1985, Tsamis2006, Youssef, Higuchi2014}. 
However our style of computation is a bit new. We compute 
by isolating the transverse part from longitudinal one. This give 
us advantage to study the physical transverse part for both the 
massive and massless vectors more carefully.
Using a common methodology of computation for both the 
massive and massless fields allows us to make comparison 
at each level of computation for both the massive and massless 
fields, and gives us confidence in our results due to 
consistency of our findings. In the massless limit the massive 
transverse part is seen to smoothly go over to the massless 
transverse part. The longitudinal part doesn't share this beauty. 
It only means that the problematic mass-divergence is only 
in the longitudinal sector of the theory and therefore is unphysical.

The outline of the paper is the following. In section \ref{Bi-tensor}
we cover some basics on bi-tensors. In section \ref{Maxsym},
we study bi-tensors on maximally symmetric space times, in 
particular deSitter. In section \ref{massvec} we study the 
massive vector fields, and compute the propagator of the theory
on a deSitter background. In section \ref{massless_sp1} we study the
case of massless vector fields. In section \ref{CPTW} we make a 
comparison of our results with the ones of the past literature. 
We conclude in section \ref{conc}
with summary, conclusion from the results and an outlook. 

\section{Bi-tensors}
\label{Bi-tensor}

In this section we will study Bi-Tensors in general and their properties.
We start by considering the motion of a particle on curved space-time from 
one point $x^{\prime \mu}$ to another point $x^\mu$. As it is a 
free particle therefore its Lagrangian is given by,
\beq
\label{eq:LagFreepart}
L = \frac{1}{2} g_\mn \frac{{\rm d} x^\mu}{{\rm d} \tau} 
\frac{{\rm d}x^\nu}{{\rm d}\tau} \, ,
\eeq
where $\tau$ is the proper time. One can work out the momentum 
of the particle to be,
\beq
\label{eq:momFreepart}
p_\mu = \frac{ \pt L}{\pt \dot{x}^\mu} = g_\mn \dot{x}^\nu = \dot{x}_\mu \, ,
\eeq
where `dot' indicates derivative with respect to local time.
Using the expression for the momentum it is easy to work out the 
Hamiltonian of the system,
\beq
\label{eq:HamFreepart}
H = p_\mu \dot{x}^\mu - L = L
\eeq
For the particle moving from position $x^{\prime \mu}$ to 
position $x^\mu$, the action for the this will be given by,
\beq
\label{eq:actFreepart}
S[x, \tau; x^\prime, \tau^\prime]
= \int_{\tau^\prime}^\tau L {\rm d} \tau
= \frac{\sg(x, x^\prime)}{\tau - \tau^\prime} \, ,
\eeq
where $\sg(x,x^\prime)$ is half the square of the geodesic distance between
$x$ and $x^\prime$. Its called \textit{geodetic interval}. After having written 
the action, we can use the Hamilton-Jacobi equation at the two end points
to obtain some identities. Hamilton-Jacobi equations at the 
end point $x$ are the following,
\beq
\label{eq:HamJacob_t}
\frac{\pt S}{\pt x^\mu} = p_\mu  \, , 
\hspace{5mm}
\frac{\pt S}{\pt \tau} + H = 0 \, .
\eeq
The first equation in eq. (\ref{eq:HamJacob_t}) gives,
\beq
\label{eq:HamJacob_t1}
p_\mu = \dot{x}_\mu = \frac{\sg_\mu}{\tau - \tau^\prime} \, ,
\eeq
where $\sg_\mu$ is the covariant derivative of $\sg$ at the 
point $x$. By making use of eq. (\ref{eq:HamJacob_t1}) we obtain 
an expression for the Hamiltonian of the system in terms of 
$\sg$. This is given by,
\beq
\label{eq:HinSigma}
H = L = \frac{\sg_\mu \sg^\mu}{2(\tau - \tau^\prime)^2} \, .
\eeq
Partial derivative of the action with respect to $\tau$ is given by,
\beq
\label{eq:dSbydt}
\frac{\pt S}{\pt \tau} = 
- \frac{\sg}{(\tau - \tau^\prime)^2} \, .
\eeq
This can be used in the second equation of 
(\ref{eq:HamJacob_t}) to get,
\beq
\label{eq:HamJacon_t2}
\frac{\pt S}{\pt \tau} + H
= \frac{1}{(\tau - \tau^\prime)^2}
\biggl[
\frac{1}{2} \sg_\mu \sg^\mu - \sg
\biggr] =0 \, .
\eeq
This tells an important relation that,
\beq
\label{eq:sigma_x}
\sg_\mu \sg^\mu = 2 \sg \, .
\eeq
The Hamilton-Jacobi equations at the other point $\tau^\prime$
are a bit different and are given by following,
\beq
\label{eq:HamJacob_tp}
\frac{\pt S}{\pt x^{\prime\mu}} = - p^\prime_\mu  \, , 
\hspace{5mm}
\frac{\pt S}{\pt \tau^\prime} - H = 0 \, .
\eeq
The first equation of eq. (\ref{eq:HamJacob_tp}) gives,
\beq
\label{eq:HamJacob_tp1}
p^\prime_\mu = \dot{x}^\prime_{\mu^\prime}
= -\frac{\sg_{\mu^\prime}}{\tau - \tau^\prime} \, .
\eeq
From eq. (\ref{eq:HamJacob_t1} and \ref{eq:HamJacob_tp1}) 
we notice that $\sg_\mu$ and $\sg_{\mu^\prime}$ are 
vectors of equal length but with opposite sign. By making use of
the second equation in eq. (\ref{eq:HamJacob_tp}), we get the 
following,
\bea
&&
\frac{\pt S}{\pt \tau^\prime} - H
= \frac{1}{(\tau-\tau^\prime)^2} \biggl[
\sg - \frac{1}{2} \sg_{\mu^\prime} \sg^{\mu^\prime}
\biggr] =0 \, ,
\notag \\
&&
\label{eq:sigma_xp}
\sg_{\mu^\prime} \sg^{\mu^\prime} = 2 \sg \, .
\eea
Although the Hamilton-Jacobi equations at the two points $x$ and 
$x^\prime$ are different, but the relation given in eq. (\ref{eq:sigma_x}
and \ref{eq:sigma_xp}) turns out to be of same form. At this point we 
define a bi-vector $g_\mnp (x, x^\prime)$. 
This is a parallel-displacement bi-vector whose purpose is to 
parallel transport a generic vector from point $x^\prime$ to $x$ and
vice-versa. It is defined by the property,
\beq
\label{eq:bivec_def}
\sg^\tau \nb_\tau g_\mnp = 0 \, .
\eeq
An arbitrary vector $V^\mu$ is said to be parallel transported 
if it satisfies the equation,
\beq
\label{eq:PT_V}
\dot{x}^\mu \nb_\mu V^\rho =0 \, .
\eeq
Using eq. (\ref{eq:HamJacob_t1}) we realise that $\dot{x}^\mu$
can be expressed in terms of $\sg_\mu$ as $\dot{x}^\mu = (\tau-\tau^\prime)
\sg^\mu$. This allow us to re-express the condition of parallel 
transport of vector $V^\mu$, eq. (\ref{eq:PT_V}) 
as $\sg^\mu \nb_\mu V^\rho =0$.
Now if we define the bi-vector $g_\mnp$ such that
$V^\mu = g_\mnp V^{\nu^\prime}$, then it is noticed
that in order for $V^\mu$ to satisfy parallel transport equation
(\ref{eq:PT_V}), $g_\mnp$ must satisfy the condition
stated in eq. (\ref{eq:bivec_def}).

Using the definition of bi-vector $g_\mnp$
it is easy to verify that it satisfies the following properties,
\bea
\label{eq:bivec_prop}
&&
g_\mnp \sg^{\nu^\prime} = - \sg_\mu \, , 
\hspace{5mm}
g_\mnp \sg^\mu = -\sg_{\nu^\prime} \, ,
\notag \\
&&
g_{\mu\rho^\prime} g_{\nu}{}^{\rho^\prime} = g_\mn \, ,
\hspace{5mm}
g_{\rho \mu^\prime} g^\rho{}_{\nu^\prime} = g_{\mu^\prime \nu^\prime} \, ,
\notag \\
&&
(\det \, g^{\nu \rho^\prime}) = \frac{1}{(\det \,  g_{\nu\rho^\prime})} \, ,
\,\,\,\, {\rm and}
\notag \\
&&
\det g_\mnp = \sqrt{g} \sqrt{g^\prime} \, .
\eea
These properties can be proved easily. The first line can be proved
by writing $\sg^\mu$ and $\sg^{\nu^\prime}$ in terms of 
$\dot{x}^\mu$ and $\dot{x}^{\nu^\prime}$ respectively (which are 
tangent vectors at the point $x$ and $x^\prime$ respectively). Using the 
fact that $g_\mnp$ parallel transports $\dot{x}^\mu$ and
$\dot{x}^{\nu^\prime}$, it is easy to prove the first line. Second 
line can be proved by contracting both sides of the equality by
$\sg^\mu \sg^\nu$ and $\sg^{\mu\prime} \sg^{\nu\prime}$ respectively.
On making use of the property $\sg_\mu\sg^\mu = 2 \sg$ and 
$\sg_{\mu^\prime} \sg^{\mu^\prime} = 2\sg$, second line follows.
The proof of the third line is a bit involved and we do it below.
First we obtain the bi-vector with indices raised. This can be
done by making use of the metric at $x$ and $x^\prime$. 
\beq
\label{eq:bivec_raised}
g^\mnp = g^{\mu\rho} g^{\nu^\prime \al^\prime} g_{\rho\al^\prime} \, .
\eeq
Taking determinant on both sides gives,
\beq
\label{eq:detrel}
\frac{1}{g} \cdot (\det  g_{\rho\al^\prime}) \cdot \frac{1}{g^\prime}
= (\det g^\mnp) \, .
\eeq
Similar to $g_\mnp$, $g^\mnp$ is also a bi-vector
which acts as a parallel displacement vector and satisfies the identities 
similar to the first line of eq. (\ref{eq:bivec_prop}). This can be derived 
from the property of $g_\mnp$. 
\bea
\label{eq:bivec_raised1}
g^\mnp \sg_{\nu\prime} 
&=& g^{\mu\al} g^{\nu\prime \bt^\prime} g_{\al\bt^\prime} \sg_{\nu^\prime} \, ,
\notag \\
&=& g^{\mu\al} g_{\al\bt^\prime} \sg^{\bt^\prime}
= g^{\mu\al} (-\sg_\mu) = -\sg^\al \, .
\eea
Now by taking the determinant of both sides of
\beq
\label{eq:bivec_inv}
g_{\mu\rho^\prime} g^{\nu\rho^\prime} = \de_\mu^\nu \, ,
\eeq
we notice that the determinant of the bi-vector with the raised 
indices is reciprocal of the determinant of the bi-vector $g_\mnp$.
Plugging this information in eq. (\ref{eq:detrel}), we prove the 
third line of eq. (\ref{eq:bivec_prop}). 

From these properties and relations one can obtain some 
more useful identities. By taking derivative of 
$\sg_\mu \sg^\mu = 2 \sg$ and $\sg_{\mu^\prime} \sg^{\mu\prime} =2\sg$ 
one can obtain certain nice identities.
\bea
\label{eq:sg_der1}
&&
\sg^\mu \sg_\mn = \sg_\nu \, , \hspace{5mm}
\sg^\mu \sg_\mnp = \sg_{\nu^\prime} \, ,
\notag \\
&&
\sg^{\mu^\prime} \sg_{\mu^\prime \nu} = \sg_\nu \, ,
\hspace{5mm}
\sg^{\mu^\prime} \sg_{\mu^\prime \nu^\prime} = \sg_{\nu^\prime} \, .
\eea
By taking derivatives of $g_\mnp \sg^{\nu^\prime} = -\sg_\mu$
and $g_\mnp \sg^{\mu} = -\sg_{\nu^\prime}$, we obtain 
the following relations,
\bea
\label{eq:bivec_der1}
&&
g_{\mu\nu^\prime; \nu} \sg^{\nu^\prime} + g_\mnp \sg^{\nu^\prime}{}_\nu
= - \sg_\mn \, ,
\notag \\
&&
g_{\mu\nu^\prime; \mu^\prime} \sg^{\nu^\prime} 
+ g_\mnp \sg^{\nu^\prime}{}_{\mu^\prime}
= - \sg_{\mu\mu^\prime} \, , \notag \\
&&
g_{\mu\nu^\prime; \nu} \sg^\mu + g_\mnp \sg^\mu{}_\nu
= - \sg_{\nu^\prime\nu} \, ,\notag\\
&&
g_{\mu\nu^\prime; \mu^\prime} \sg^\mu + g_\mnp \sg^\mu{}_{\mu^\prime}
= - \sg_{\nu^\prime\mu^\prime} \, .
\eea
In the same way one can take more derivatives and compute further identities
\footnote{These will become relevant when one is computing the heat-kernel 
coefficients of the differential operators on curved backgrounds}.

\section{Maximally-Symmetric Spacetimes}
\label{Maxsym}

Maximally symmetric space-time are manifolds in arbitrary dimensions 
having the largest number of symmetries possible. They look the same
in all directions. The simplest example for this is the flat space-time and next in
simplicity is the $d$-sphere or hyperboloid (in Reimannian spaces).
Leaving apart space-time with zero curvature (which is flat space-time),
one is left with space-time having constant nonzero curvature. These 
include $d$-sphere and hyperboloid in Euclidean spaces
while DeSitter and Anti-DeSitter in Pseudo-Riemanian 
spaces (which have constant positive or negative curvatures). 
These kind of spaces have $d(d+1)/2$ independent killing vectors.

In a maximally symmetric background any arbitrary bi-scalar is just
a function of the bi-scalar given by the geodetic distance $\sg(x, x^\prime)$,
any arbitrary bi-vector $S_\mnp$ can be written as a linear combination of 
parallel displacement bi-vector $g_\mnp$ and 
$\sg_\mu \sg_{\nu^\prime}$ with coefficients which are functions 
of bi-scalar $\sg(x, x^\prime)$. Similarly higher rank arbitrary bi-tensors 
$T_{\mu\nu\rho \cdots}{}^{\mu^\prime \nu^\prime \rho^\prime \cdots}$
can be written as a linear combination of bi-tensors constructed using 
$g_\mnp$, $\sg_\mu$ and $\sg_{\mu^\prime}$,
with coefficients being functions of $\sg(x, x^\prime)$. This property is only
true for maximally symmetric spaces and no longer holds for 
arbitrary backgrounds. This was first discussed in \cite{Katz, Peters}
(check \cite{AllenJacob1985} also).

For a maximally symmetric space-time one can write $\sg_{\mu\nu}$
as a linear combination of $g_\mn$ and $\sg_\mu \sg_\nu$, with 
coefficients as functions of the geodetic distance $\sg(x, x^\prime)$.
Similarly one can write $\sg_\mnp$ as a a linear combination
of $g_\mnp$ and $\sg_\mu\sg_{\nu^\prime}$.
\bea
\label{eq:smn}
&&
\sg_\mn = A(\sg) g_\mn + B(\sg) \sg_\mu \sg_\nu \, , \\
&&
\label{eq:smnp}
\sg_\mnp = C(\sg) g_\mnp
+ D(\sg) \sg_\mu \sg_\nu^\prime \, .
\eea
It is easy to solve for the coefficients $A$, $B$, $C$ and $D$.
By contracting the eq. (\ref{eq:smn}) first with $g^\mn$
and then with $\sg_\mu$, we get the following two equations:
\beq
\label{eq:smn_ABsolve}
\Box \sg = d A(\sg) + 2 \sg B(\sg) \, \hspace{5mm}
1 = A(\sg) + 2\sg B(\sg) \, .
\eeq
From the second equation in eq. (\ref{eq:smn_ABsolve}) 
one can solve for $B$ in terms of $A$, while from the 
first equation one can work out $A$ in terms of $\Box \sg$.
Finally one can write $\sg_\mn$ in the following way,
\bea
\label{eq:smn_soln}
&&
\sg_\mn = A(\sg) \biggl[
g_\mn - \frac{1}{2\sg} \sg_\mu \sg_\nu
\biggr] + \frac{1}{2\sg} \sg_\mu \sg_\nu \, , 
\notag \\
{\rm with}\,
&&
A(\sg) = \frac{\Box \sg - 1}{d-1} \, ,
\eea
where $\Box\sg$ is to be determined later and depends on 
the background space-time and its signature.
Similarly contracting eq. (\ref{eq:smnp}) with $\sg^\mu$,
gives $D(\sg)$ in terms of $C(\sg)$, plugging which we get the
following:
\beq
\label{eq:sgmnp_C}
\sg_\mnp = C(\sg) \biggl[
g_\mnp + \frac{1}{2\sg} \sg_\mu \sg_{\nu^\prime}
\biggr] + \frac{1}{2\sg} \sg_\mu \sg_{\nu^\prime} \, .
\eeq 
However solving for $C$ is a bit involved. We apply $\sg^\al \nb_\al$
on both sides of eq. (\ref{eq:sgmnp_C}). The LHS becomes 
$\sg^\al \nb_\al \sg_\mnp$. As $\sg_{\nu^\prime}$ is a scalar
at $x$, therefore the indices $\al$ and $\mu$ commute, 
thereby implying 
\beq
\label{eq:dercommute}
\sg^\al\nb_\al \sg_\mnp = 
\sg^\al \nb_\mu \sg_{\al\nu^\prime} \, .
\eeq
In order to compute the expression for $\sg^\al \nb_\mu \sg_{\al\nu^\prime}$, 
we apply the derivative $\nb_{\nu^\prime}$ on the identity 
$\sg^\al \sg_{\al\mu} = \sg_\mu$. This gives,
\beq
\label{eq:dersg_4}
\sg^\al \sg_{\mu\al\nu^\prime}
= \sg_\mnp - \sg^\al{}_\mu \sg_{\al\nu^\prime} \, .
\eeq
Plugging the expressions from eq. (\ref{eq:smn_soln} and
\ref{eq:sgmnp_C}), we get an expression for 
$\sg^\al \nb_\mu \sg_{\al\nu^\prime}$ to be,
\beq
\label{eq:sg4Csolve}
\sg^\al \sg_{\al\mu\nu^\prime}
= C(1-A) \biggl[g_\mnp 
+ \frac{1}{2\sg} \sg_\mu \sg_{\nu^\prime}
\biggr] \, .
\eeq
This is also the expression obtained by action of $\sg^\al \nb_\al$
on the LHS of eq. (\ref{eq:sgmnp_C}). Action of $\sg^\al \nb_\al$
on the RHS of eq. (\ref{eq:sgmnp_C}) gives the following,
\beq
\label{eq:smnp_Cder}
\sg^\al \nb_\al \sg_\mnp
= 2\sg C^\prime(\sg) \biggl[
g_\mnp + \frac{1}{2\sg} \sg_\mu \sg_{\nu^\prime}
\biggr] \, ,
\eeq
where $C^\prime(\sg)$ means derivative of $C$ with respect to $\sg$.
Comparing the result in eq. (\ref{eq:smnp_Cder}) with the result 
from eq. (\ref{eq:sg4Csolve}) we finally get an equation for $C$. 
This is given by,
\beq
\label{eq:smnp_Ceq}
C^\prime = \frac{C(1-A)}{2\sg} \, .
\eeq
Being a first order differential equation, it can be integrated 
easily to give the following solution,
\beq
\label{eq:Csoln}
C = {\rm const.} \, \exp[ 
\int \frac{C(1-A)}{2\sg} {\rm d}\sg] \, .
\eeq
The constant of integration is determined by comparing with the 
short distance (flat space) behaviour of $C$. To do so it is
first convenient to study the covariant derivative of the 
parallel displacement bi-vector $g_{\al\bt^\prime}$. 
On a maximally symmetric space 
$g_{\al\bt^\prime; \mu}$ can be written in the following way,
\bea
\label{eq:bivec_der}
g_{\al\bt^\prime; \mu} 
&=& E g_{\al\mu} \sg_{\bt^\prime}
+ F g_{\mu\bt^\prime} \sg_\al
+ G g_{\al\bt^\prime} \sg_\mu
\notag \\
&+& H \sg_\al \sg_{\bt^\prime} \sg_\mu \, .
\eea
As $g_{\al\bt^\prime}$ satisfies the eq. (\ref{eq:bivec_def}),
therefore contracting eq. (\ref{eq:bivec_der}) with $\sg^\mu$
we get the following condition on the coefficients,
\beq
0 = E - F + 2\sg H \, , \hspace{5mm}
2\sg G =0 \, .
\eeq
Contraction of LHS of eq. (\ref{eq:bivec_der}) with $\sg^\al$, 
and making use of the identities stated in 
eq. (\ref{eq:bivec_der1}), (\ref{eq:bivec_prop}) and 
(\ref{eq:sigma_x}) gives us an expression where the 
solution from eq. (\ref{eq:smn_soln}) and (\ref{eq:sgmnp_C})
can be plugged to obtain the following,
\beq
\label{eq:sg_bivecder1}
\sg^\al g_{\al\bt^\prime; \mu}
= -(A+C) \biggl[
g_{\mu\bt^\prime} + \frac{1}{2\sg} \sg_\mu \sg_{\bt^\prime}
\biggr] \, .
\eeq
Similarly LHS of eq. (\ref{eq:bivec_der}) can be contracted 
with $\sg^{\bt^\prime}$. Again by making use of the 
identities stated in eq. (\ref{eq:bivec_der1}), (\ref{eq:bivec_prop}) and 
(\ref{eq:sigma_x}) gives us an expression where the 
solution from eq. (\ref{eq:smn_soln}) and (\ref{eq:sgmnp_C})
can be plugged to obtain the following,
\beq
\label{eq:sg_bivecder2}
\sg^{\bt^\prime} g_{\al\bt^\prime; \mu}
= -(A+C) \biggl[
g_{\mu\al} + \frac{1}{2\sg} \sg_\mu \sg_\al
\biggr] \, .
\eeq
The results from eq. (\ref{eq:sg_bivecder1}) and 
(\ref{eq:sg_bivecder2}) can be compared by contracting 
RHS of eq. (\ref{eq:bivec_der}) with $\sg^\al$ and 
$\sg^{\bt^\prime}$ respectively, thereby giving the 
following relations respectively,
\bea
\label{eq:reln_EFGH1}
&&
E - G + 2\sg H = - \frac{A+C}{2\sg} \, ,
\,
F = -\frac{A+C}{2\sg} \, , \\
&&
\label{eq:reln_EFGH2}
E = -\frac{A+C}{2\sg} \, , 
\,
-F - G + 2\sg H = -(A+C) \, .
\eea
Solving these relations simultaneously we get the expression for the
derivative of bi-vector $g_{\al\bt^\prime}$ to be,
\beq
\label{eq:bivec_derexp}
g_{\al\bt^\prime;\mu} = - \frac{A+C}{2\sg}
(g_{\mu\al} \sg_{\bt^\prime} + g_{\mu\bt^\prime} \sg_\al) \, .
\eeq
In flat space-time or at short distances $A+C$ must vanish. Hence the
constant of integration in eq. (\ref{eq:Csoln}) is determined 
by requiring
\beq
\label{eq:flatcond}
\lim_{\sg \to 0} [A(\sg) + C(\sg) ] =0 \, .
\eeq
The bi-scalar $\Box\sg(x, x^\prime)$ which appears in the expression of $A$ is
again maximally symmetric. Therefore it must depend only on 
geodetic distance between $x$ and $x^\prime$. To obtain an exact 
expression of this requires knowledge of the kind of maximally symmetric 
background for which it is computed and its signature. 
We will evaluate here $\Box\sg$ for two special different kind of 
euclidean maximally symmetric space-time: 
zero curvature flat $\mathbb{R}^n$ and positive curvature $S^n$. 
All of these spaces have constant curvature.

In order to evaluate $\Box \sg$ in $\mathbb{R}^n$, we go to the spherical 
co-ordinates and centre the co-ordinate system around $x^\prime$. The 
metric in Euclidean space will be given by, $ds^2 = dr^2 + r^2 d\OM_{d-1}^2$.
The geodetic distance will only be a function of $r$ co-ordinate. As $\sg$ is half of the
square of the geodesic distance between two points, therefore in flat space-time,
for the co-ordinate system described above $\sg = r^2/2$. In general 
the $\Box$ operator acting on a scalar has the following expression,
\beq
\label{eq:box_scalar}
\Box \phi = \frac{1}{\sqrt{g}} \pt_\mu
[g^\mn \sqrt{g} \pt_\nu \phi] \, ,
\eeq
where $g$ is the determinant of the metric $g_\mn$
and $\phi$ is some arbitrary scalar. For flat space-time, as
$\sg$ is just a function of $r$, therefore in the $\Box$
operator only $g^{rr}$ contributes, thereby implying,
\beq
\label{eq:Boxonsg1}
\Box \sg = \frac{1}{r^{d-1}} \frac{{\rm d}}{{\rm d} r} 
r^{d-1} \frac{{\rm d}}{{\rm d} r} \sg = d \, .
\eeq
This means that for flat euclidean case $A=1$ and
$C=-1$. Now we proceed to work out the case with positive 
curvature $S^n$. For a $d$-sphere with the co-ordinate 
centred around $x^\prime$, the metric is given by
$ds^2 =  r^2 d\ta^2 + r^2 \sin^2 \ta d \OM^2_{d-1}$, where 
$r$ is the fixed radius of the sphere and is related to 
Ricci scalar $R=d(d-1)/r^2$. On a $d$-sphere 
for the above metric, $\sg = r^2 \ta^2 /2$.
Again the $\Box$ operator for this metric will contain in general 
lot of terms, however when $\Box$ acts on $\sg$, it will only
pick the contribution corresponding to $g^{\ta\ta}$ from
eq. (\ref{eq:box_scalar}). This will imply that,
\bea
\label{eq:Boxonsg2}
\Box \sg 
&=& \frac{1}{r^2} \frac{1}{\sin^{d-1}\ta}
\frac{{\rm d}}{{\rm d} \ta} \sin^{d-1}\ta 
\frac{{\rm d}}{{\rm d} \ta} \sg
= 1+ (d-1) \ta \cot \ta
\notag \\
&=& 1+(d-1) \sqrt{\frac{2\sg R}{d(d-1)}}
\cot \sqrt{\frac{2\sg R}{d(d-1)}} \, .
\eea
Once $\Box \sg$ on DeSitter is found, it can be used to compute the 
expression for $A$ and $C$ using eq. (\ref{eq:smn_soln} and 
\ref{eq:Csoln}). The constant in eq. (\ref{eq:Csoln}) is determined 
by requiring that in the flat limit ($\sg\to0$), $(A+C) \to 0$.
This would determine $C$ completely. The values of $A$ and 
$C$ on a euclidean DeSitter are given by,
\beq
\label{eq:ACvalds}
A = \sqrt{\frac{2\sg R}{d(d-1)}} \cot \sqrt{\frac{2\sg R}{d(d-1)}}
\, , 
\hspace{5mm}
C = -\sqrt{\frac{2\sg R}{d(d-1)}} \csc \sqrt{\frac{2\sg R}{d(d-1)}} \, .
\eeq

\section{Two-point function for Massive Vector}
\label{massvec}

In this section we compute the two-point function for the massive 
vector fields on the deSitter background. The action for the 
theory in euclidean space is given by,
\beq
\label{eq:act_massvec}
S = \int {\rm d}^dx \sqrt{g} 
\biggl[
\frac{1}{4} F_\mn F^\mn + \frac{1}{2} m^2 A_\mu A^\mu 
\biggr] \, ,  
\eeq
where $F_\mn = \nb_\mu A_\nu - \nb_\nu A_\mu$ is the 
field strength tensor antisymmetric in two indices, while 
$A_\mu$ is the spin-1 vector field. The action given in 
eq. (\ref{eq:act_massvec}) is also the action for the
Proca theory for massive spin-1 particle.
It also arise in standard model of particle physics
when there is spontaneous electroweak symmetry breaking 
via Higgs mechanism, thereby giving mass to the vector bosons.
We consider the euclidean, source dependent path-integral for this
theory, which is given by following,
\beq
\label{eq:MV_pathint}
Z[J] = \int {\cal D} A_\mu 
\exp\biggl[
- \int {\rm d}^dx \sqrt{g} 
\biggl(
\frac{1}{4} F_\mn F^\mn 
+ \frac{1}{2} m^2 A_\mu A^\mu 
\biggr)
- \int {\rm d}^dx \sqrt{g} J_\mu A^\mu
\biggr] \, ,
\eeq
where $J_\mu$ is the source. This is the generating functional for the 
various correlation functions. From this we can define the 
connected Green functional $W[J] = - \ln Z[J]$, consisting only of 
connected Feynman diagrams. The action in eq. (\ref{eq:act_massvec})
being quadratic, allows us to perform the gaussian integral over 
the field in the path-integral eq. (\ref{eq:MV_pathint}), thereby 
giving an expression for $W[J]$ to be,
\beq
\label{eq:WJ_MV}
W[J] = \frac{1}{2} \int {\rm d}^dx \sqrt{g} J_\mu \left(\D_F^{-1} \right)^\mn
J_\nu + \frac{1}{2} {\rm Tr} \ln \D_F^\mn \, ,
\eeq 
where
\beq
\label{eq:Hessian}
\D_F^\mn = \left(-\Box + \frac{R}{d} + m^2 \right)g^\mn 
+ \nb^\mu \nb^\nu \, .
\eeq
From the connected green's functional, we obtain the source 
dependent expectation value of the vector field $A_\mu$,
denoted by $\bar{A}_\mu$. From the expression of 
$W[J]$ given in eq. (\ref{eq:WJ_MV}), it is easy to compute 
$\bar{A}_\mu = \left(\D_F^{-1} \right)_\mu{}^\nu J_\nu$. 
Defining the generating functional for the one-particle 
irreducible (1PI) graphs in the usual manner, we obtain 
Effective action for the theory to be,
\beq
\label{eq:1PI_EAdef}
\G[\bar{A}_\mu] 
= W[J] - \int {\rm d}^dx \sqrt{g} J_\mu \bar{A}^\mu
= -\frac{1}{2} \int {\rm d}^dx \sqrt{g} \bar{A}_\mu \D_F^\mn \bar{A}_\nu
+ \frac{1}{2} {\rm Tr} \ln \D_F^\mn \, .
\eeq
Then the derivative of $\G$ with respect to $\bar{A}_\mu$
is $-J^\mu$. In general the generating functionals $W$ and $\G$
have some interesting identities, which relate various correlation 
functions to each other. The most basic of this identity is,
\bea
\label{eq:GW_iden_func}
&&
\int {\rm d}^dy \sqrt{g(y)} \biggl(
\frac{1}{\sqrt{g(x)}} \frac{1}{\sqrt{g(y)}}
\frac{\de^2 \G}{\de \bar{A}_\mu(x) \de \bar{A}_\rho(y)}
\biggr)
\biggl(
\frac{1}{\sqrt{g(y)}} \frac{1}{\sqrt{g(x^\prime)}}
\frac{\de^2 W}{\de J^\rho(y) \de J_{\nu^\prime}(x^\prime)}
\biggr)
\notag \\
&&
= - \frac{\de_\mu{}^{\nu^\prime} \de(x-x^\prime)}{\sqrt{g(x)}} \, .
\eea
Given that we had an action for non-interacting theory, where $W$ and $\G$ 
have simple expressions, it is easy to compute the double derivatives
of either of them which are given by following,
\bea
\label{eq:2derW}
&&
\frac{1}{\sqrt{g}} \frac{1}{\sqrt{g^\prime}}\frac{\de^2 W}{\de J_\mu(x) \de J_{\nu^\prime}(x^\prime)}
=\langle A_\mu(x)  A_{\nu^\prime}(x^\prime) \rangle 
= \frac{1}{\sqrt{g^\prime}} \left(\D_F^{-1}\right)^\mnp \de(x-x^\prime) 
= G^\mnp(x, x^\prime) \, ,
\\
\label{eq:2derGamma}
&&
\frac{1}{\sqrt{g}} \frac{1}{\sqrt{g^\prime}} 
\frac{\de^2 \G}{\de \bar{A}_\mu(x) \de \bar{A}_{\nu^\prime}(x^\prime)}
= - \frac{1}{\sqrt{g^\prime}} \D_F^\mnp\de(x-x^\prime) \, .
\eea
Plugging this in the identity eq. (\ref{eq:GW_iden_func}) and doing 
integration by parts fetches us the equation for the two-point 
function $G_\mnp(x, x^\prime)$. This is given by,
\beq
\label{eq:GRMV_curved}
\biggl[
\left(-\Box + \frac{R}{d} + m^2 \right)\de_\mu{}^\al 
+ \nb_\mu \nb^\al 
\biggr] G_{\al\nu^\prime}
= \frac{g_\mnp \de(x-x^\prime)}{\sqrt{g^\prime}} \, .
\eeq
In principle by solving this equation for the $G_\mnp$ we get 
the green's function for the massive vector fields on the 
deSitter background. However, here we will adopt a different 
strategy of computing the green's function. We will solve the
above equation for the non-coincident points. This has the advantage of
simplicity. Then we require that at short distance the green's function 
on the curved background should match the green's function of the
fields on the flat space-time. This is equivalent to putting the 
appropriate boundary condition or solving for the green's function
with the delta-function on the RHS of eq. (\ref{eq:GRMV_curved}). 
In the following we will solve the Green's function for the 
non-coincident points, where the appropriate boundary condition 
is applied by requiring the green's function on curved background 
to match the flat space-time structure at short distance. The equation
satisfied by the green's function for non-coincident points is,
\beq
\label{eq:EQM_MV}
\biggl[\left(-\Box + \frac{R}{d} + m^2 \right)g^\mn 
+ \nb^\mu \nb^\nu 
\biggr] \langle A_\nu A_{\nu^\prime} \rangle =0 \, .
\eeq
Using this equation we can compute the expression for 
$\nb^\mu \langle F_\mn A_{\nu^\prime} \rangle$ by doing 
some manipulations. This will yield,
\beq
\label{eq:nabF}
\nb^\mu \langle F_\mn A_{\nu^\prime} \rangle
= m^2 \langle A_\nu(x)  A_{\nu^\prime}(x^\prime) \rangle \, .
\footnote{
More precisely there is a $\de$-function term on the RHS also,
which is given by $-g_{\nu\nu^\prime} \de(x-x^\prime)/\sqrt{g^\prime}$.
But as we are interested in solving for the non-coincident points, 
therefore this terms vanishes. However this then later translates over the 
condition $m^2 \nb^\nu \langle A_\nu A_{\nu^\prime} \rangle
= \nb^\nu [g_{\nu\nu^\prime} \de(x-x^\prime)/\sqrt{g^\prime}]$,
instead of eq. (\ref{eq:supconst_MV}). 
}
\eeq
It should be noticed that contracting eq. (\ref{eq:nabF}) with 
$\nb^\nu$ and using the anti-symmetry property of $F_\mn$, 
we get an additional constraint
\beq
\label{eq:supconst_MV}
m^2 \nb^\nu \langle A_\nu(x)  A_{\nu^\prime}(x^\prime) \rangle =0 \, .
\eeq
As $m^2\neq0$, therefore this will imply an additional constraint 
on the green's function. In the case when $m^2=0$ there is a gauge 
symmetry in the system, and then this constraint acts as 
Landau gauge-fixing condition. Massive vector studies done
in the past  \cite{AllenJacob1985, Tsamis2006} have made use of this 
constraint in their analysis. While the former computes the 
propagator of the Proca theory, the later computes the propagator of
the Stueckelberg field in the landau gauge. Here we compute the 
propagator of the massive theory in an alternative way 
via analysing the $\langle F_\mn F_{\mu^\prime\nu^\prime} \rangle$ 
correlator. We isolate the longitudinal piece of the 
propagator which is seen to diverge as $1/m^2$ in the massless limit.

Now we will make use of the eq. (\ref{eq:EQM_MV}, \ref{eq:nabF} and 
\ref{eq:supconst_MV}) to solve for the full propagator of the theory. 
At this point we decompose the green's function 
$G_\mnp$ in to transverse and longitudinal parts as,
\beq
\label{eq:GTV_decomp}
G_\mnp(x, x^\prime) 
= G^T_\mnp + \nb_\mu \nb_{\nu^\prime} G \, ,
\eeq
where $G^T_\mnp$ satisfies the transversality condition
\beq
\label{eq:trans_G}
\nb^{\mu} G^T_\mnp = 
\nb^{\nu^\prime} G^T_\mnp =0 \, .
\eeq
This decomposition can be justified by splitting the 
vector field $A_\mu$ in to transverse and longitudinal part 
as $A_\mu = A_\mu^T + \nb_\mu a$, where $a$ is the longitudinal 
part of the vector field, while the transverse part satisfies 
$\nb^\mu A_\mu^T=0$. This will imply,
\beq
\label{eq:GRMV_fieldD}
G_\mnp^T = \langle A_\mu^T A_{\nu^\prime}^T \rangle \, ,
\hspace{5mm}
G(x, x^\prime) = \langle a(x) a(x^\prime) \rangle \, .
\eeq
This decomposition can be plugged in (\ref{eq:EQM_MV}) 
and use eq. (\ref{eq:supconst_MV}) (for $m^2\neq0$), 
it is easy to note the following,
\beq
\label{eq:TLreln_EQM_MV}
\biggl[
-\Box + \frac{R}{d} + m^2 
\biggr] \biggl[G^T_\mnp +
\nb_\mu \nb_{\nu^\prime} G (x, x^\prime)
\biggr] = 0\, .
\eeq
We start by studying the transverse part first.

\subsection{Transverse part}
\label{TRGMV}

On a maximally symmetric background it is possible to write the general 
structure of the transverse green function in terms of parallel displacement 
bi-vector $g_{\al\bt^\prime}$ and $\sg_\al \sg_{\bt^\prime}$. This freedom
is not available on an arbitrary background, but only on the maximally 
symmetric background this privilege is enjoyed, which allows
us to write the bi-vector $G^T_\mnp$ as,
\beq
\label{eq:TV_GMV}
G^T_\mnp
= \al(\sg) g_\mnp
+ \bt(\sg) \sg_\mu \sg_{\nu^\prime} \, .
\eeq
At this point we consider the quantity $\langle F_\mn F^{\mu^\prime\nu^\prime}\rangle$.
This quantity is completely independent of the longitudinal part of the
gauge field and therefore depends only on the transverse part of the 
green function $G^T_\mnp$. On a maximally symmetric 
background it can be expressed as a linear combination 
of $g_{[\mu}{}^{[\mu^\prime}g_{\nu]}{}^{\nu\prime]}$
and $\sg_{[\mu}\sg^{[\mu^\prime}g_{\nu]}{}^{\nu\prime]}$, with coefficients 
being function of the bi-scalar $\sg$ as follows,
\beq
\label{eq:FF_exp}
\langle F_\mn F^{\mu^\prime\nu^\prime} \rangle
= 4 \nb_{[\mu} \nb^{[\mu^\prime} \langle A_{\nu]} A^{\nu^\prime]} \rangle
= \ta(\sg) g_{[\mu}{}^{[\mu^\prime}g_{\nu]}{}^{\nu\prime]}
+ \tau(\sg) \sg_{[\mu}\sg^{[\mu^\prime}g_{\nu]}{}^{\nu\prime]} \, .
\eeq
By plugging the expression for the green's function for the massive 
vector field from eq. (\ref{eq:TV_GMV}) in eq.(\ref{eq:FF_exp})
we obtain relations expressing $\ta$ and $\tau$ in terms of 
$\al$ and $\bt$ respectively. These relations are given by,
\beq
\label{eq:albt_MV}
\ta = 4C \left[
\al^\prime + \frac{A+C}{2\sg} \al - \bt C
\right] \, ,
\hspace{5mm}
\tau = C^{-1} \left[
\ta^\prime + \frac{A+C}{\sg} \ta
\right] \, .
\eeq
The covariant divergence of
$\langle F_\mn F^{\mu^\prime\nu^\prime} \rangle$ 
can be obtained by using eq. (\ref{eq:nabF}).
This give us a relation between the functions $\ta$ and $\tau$.
The covariant divergence is given by,
\beq
\label{eq:covDiv_FMV}
\nb^\mu \langle F_\mn F^{\mu^\prime\nu^\prime} \rangle
= 2 m^2 \nb^{[\mu^\prime} \langle A_\nu^T A^{T\nu^\prime]} \rangle \, .
\eeq
The LHS on plugging the expression for 
$\langle F_\mn F^{\mu^\prime\nu^\prime} \rangle$ and doing some 
simplification gives,
\beq
\label{eq:LHS_divFF}
\nb^\mu \langle F_\mn F^{\mu^\prime\nu^\prime} \rangle
= \biggl[
-\ta^\prime + \sg \tau^\prime
- (d-2) \frac{(A+C)}{2\sg} \ta
+ \frac{((d-2)A+2)\tau}{2}
\biggr] \sg^{[\mu^\prime} g_{\nu}{}^{\nu^\prime]} \, ,
\eeq
while the RHS of eq. (\ref{eq:covDiv_FMV}), on using the 
expression for $\ta$ from eq. (\ref{eq:albt_MV}) gives,
\beq
\label{eq:RHS_divFF}
2 m^2 \nb^{[\mu^\prime} \langle A_\nu^T A^{T\nu^\prime]} \rangle
= \frac{m^2 \ta}{2C} \sg^{[\mu^\prime} g_{\nu}{}^{\nu^\prime]} \, .
\eeq
Therefore equating the expressions obtained for LHS and RHS 
from eq. (\ref{eq:LHS_divFF}) and eq. (\ref{eq:RHS_divFF}) respectively,
and using the expression for $\tau$ in terms of $\ta$ from the 
eq. (\ref{eq:albt_MV}), gives us a differential equation for the 
function $\ta$ to be,
\beq
\label{eq:taeq_FFMV}
2\sg \ta^{\prime\prime}
+ [(d+1)A + 1] \ta^\prime - \frac{2R}{d} \ta - m^2 \ta =0\,,
\eeq
where we have used the identity $(C^2-A^2)/2\sg = R/d(d-1)$. This 
equation can be recast in a more appropriate form by doing a 
change of variable from $\sg$ to $z$, where $z$ is,
\beq
\label{eq:zdef}
z(x, x^\prime) = \cos^2 \sqrt{\frac{\sg R}{2d(d-1)}} \, .
\eeq
The equation for $\ta$ in terms of $z$ acquires the following form,
\beq
\label{eq:ta_z_FFMV}
z(1-z) \frac{{\rm d}^2 \ta}{{\rm d}z^2} 
+ \biggl[\frac{d}{2} +1 - (d+2)z \biggr] \frac{{\rm d}\ta}{{\rm d}z}
- \frac{d(d-1)}{R} \left(m^2 + \frac{2R}{d} \right) =0
\eeq
This is a second order differential equation for the hyper-geometric 
function ${}_2F_1(a_1, b_1; c_1; z)$ with the parameters given by,
\bea
\label{eq:MV_para}
&&
a_1 = \frac{1}{2} \biggl[
d+1 + \sqrt{(d-3)^2 - \frac{4d(d-1)m^2}{R}} 
\biggr] \, ,
\notag \\
&&
b_1 = \frac{1}{2} \biggl[
d+1 - \sqrt{(d-3)^2 - \frac{4d(d-1)m^2}{R}}
\biggr] \, ,
\notag \\
&&
c_1 = \frac{d}{2} +1 \, .
\eea
Equation (\ref{eq:ta_z_FFMV}) being a second order differential equation has two 
independent solutions. This differential equation is invariant under the change of variables 
from $z$ to $(1-z)$, as the parameters $a_1$, $b_1$ and $c_1$ satisfy the relation
$a_1+b_1+1=2c_1$. Such a symmetry doesn't always exist, but in this case 
this property is enjoyed. This has an advantage, as it tells 
that the eq. (\ref{eq:ta_z_FFMV}) has the following two solutions,
\beq
\label{eq:twoIndp_soln}
_2F_1(a_1, b_1; c_1; z) \hspace{2mm}
{\rm and} \hspace{2mm} _2F_1(a_1, b_1; c_1; 1-z) \, .
\eeq
The solution $_2F_1(a_1, b_1; c_1; z)$ 
is written as a power series in $z$, which is 
convergent for $\lvert z\rvert<1$. By analytic continuation
it can be extended to the rest of complex plane.
It is singular at the point $z=1$ and behaves 
like $_2F_1(a_1, b_1; c_1; z) \sim (1-z)^{-d/2}$, while it is 
regular at $z=0$ ($_2F_1(a_1, b_1; c_1; z)\to 1$). The 
behaviour near $z\to1$ corresponds to the short
distance singularity in the green's function arising when $\sg\to0$,
while the regularity at large distances ($z\to0$) is necessary 
for sensible physical theories. The other solution,
$_2F_1(a_1, b_1; c_1; 1-z)$, has a singularity at $z=0$, while 
it is regular at $z\to1$. These two solutions are then linearly 
independent as they have different singular points. 
A general solution to eq. (\ref{eq:ta_z_FFMV}) will be linear 
combination of both of them.

Generically, any solution of the differential eq. (\ref{eq:ta_z_FFMV})
being a linear combination of both $_2F_1(a_1, b_1; c_1; z)$
and $_2F_1(a_1, b_1; c_1; 1-z)$, will be singular at both 
$z=1$ and $z=0$ respectively. However boundary conditions (which come 
from physical requirements) help us in finding the right solution. 
On a Riemannian DeSitter background the values taken by
$\sg$ are between zero and $\pi^2 d(d-1)/2R$, which translates 
into a range for the values for $z$ to be $0\leq \lvert z \rvert <1$. 
This range however covers the entire Riemannian DeSitter background.

There is one parameter family of deSitter invariant fock vacuum state
\cite{Burges1984, Allen1985, Allen1987}, in each of which the Green's function can be 
determined. However there is one special vacuum called the
``Bunch-Davies'' vacuum \cite{Bunch1978}. It is the only vacuum for which
the two point function has: 1) only one singular
point at $z=1$ and is regular at $z=0$, and 2) the strength of singularity 
for $\sg \to 0$ is the same as in flat case. For these issues it appears 
to be the most reasonable vacuum to work with. In fact Green's
function for all other vacuum state can be derived from this one
\cite{GibbonHawking77}.

By making use of the first condition it is found that for deSitter 
case the function $\ta(z)$ can be written as,
\beq
\label{eq:greenFFMV}
\ta(z) = q \times {}_2F_1(a_1, b_1; c_1; z) \, ,
\eeq
where the parameter $q$ is an arbitrary constant and is determined 
by comparing the small distance limit of the $\ta(z)$ with the 
strength of singularity in flat space-time, which is the requirement 
imposed by the second condition. In the short distance limit, 
the asymptotic expression for the
hyper-geometric function ${}_2F_1(a_1, b_1; c_1; z)$ is given by,
\beq
\label{eq:sg0_hyperasp}
{}_2F_1(a_1, b_1; c_1; z) \sim 
\frac{\G(c_1) \G(a_1+b_1-c_1)}
{\G(a_1) \G(b_1)} \bigl(
1-z
\bigr)^{c_1-a_1-b_1} \, .
\eeq
This can be plugged in the expression for $\ta(z)$ 
in eq. (\ref{eq:greenFFMV}) to compute the asymptotic 
form for the function $\ta(z)$ to be,
\beq
\label{eq:FF_MV_flatlimit}
\ta_{\sg\to0} (z) \sim  \frac{q \G(c_1) \G(a_1+b_1-c_1)}
{\G(a_1) \G(b_1)} \bigl(
1-z
\bigr)^{c_1-a_1-b_1} \, .
\eeq 
This can be compared with the corresponding $\ta(z)$ 
from flat space-time case discussed in the appendix 
\ref{massvec_flat}, with the leading short distance 
singularity of $\ta(z)$ given in eq. (\ref{eq:FF_flat_lead}). 
This comparison allows us to compute the value of the 
coefficient $q$ in the eq. (\ref{eq:greenFFMV}), which is 
given by,
\beq
\label{eq:FF_MV_flatq}
q = \frac{2}{(4\pi)^{d/2}} \frac{\G(a_1)\G(b_1)}
{\G(d/2+1)} \left(\frac{R}{d(d-1)}\right)^{d/2} \, .
\eeq
Once the parameter $q$ is found, we have the knowledge of
the function $\ta(z)$. One can then work out the expression
for the function $\tau(z)$ by making use of the 
relation in eq. (\ref{eq:albt_MV}). This relation can be 
translated in the language of $z$ and is given by following,
\beq
\label{eq:tauZeq}
\tau(z) = \frac{R}{2d(d-1) (\acos\sqrt{z})^2}\biggl[
z(1-z) \frac{{\rm d}\ta}{{\rm d}z} + 2(1-z) \ta
\biggr] \, .
\eeq
Having found the behaviour of the function $\ta(z)$ 
and $\tau(z)$, we proceed to work out the transverse part 
of the green's function for massive vector field.
On a DeSitter background the structural form of the two-point 
function is given in eq. (\ref{eq:TV_GMV}). This satisfies the 
transversality constraint given in eq. (\ref{eq:trans_G}), which 
implies a relation between $\al$ and $\bt$ given by,
\beq
\label{eq:albt_relt_MV}
\al^\prime - 2\sg \bt^\prime - 2\bt
+ \frac{(d-1)(A+C)}{2\sg}\al - (d-1) A \bt =0 \, ,
\eeq
where `prime' denotes derivative with respect to the argument $\sg$.
We can use the first relation given in eq. (\ref{eq:albt_MV})
to express $\bt$ and $\bt^\prime$ in terms of $\ta$, $\ta^\prime$, 
$\al$, $\al^\prime$ and $\al^{\prime\prime}$. 
These relations are given by,
\bea
\label{eq:btinta_MV}
&&
\bt = C^{-1} \biggl[
\al^\prime + \frac{A+C}{2\sg} \al -\frac{1}{4} \ta C^{-1}
\biggr] \, ,
\notag \\
&&
\bt^{\prime}= C^{-1} \biggl[
\al^{\prime\prime} + \frac{2A+C-1}{2\sg} \al^\prime
+ \frac{(A+C)(A-C-2)}{2\sg} \al
-  \frac{\ta^\prime}{4C}
+ \frac{\ta(1-A)}{4C\sg}
\biggr] \, .
\eea
These relation can be used to obtain an equation for the variable 
$\al$ by plugging them in eq. (\ref{eq:albt_relt_MV}). After doing some
simplification we get the following equation,
\beq
\label{eq:aldiff_MV}
2\sg \al^{\prime\prime} 
+ [(d+1)A+1] \al^\prime 
- \frac{R}{d-1} \al = \frac{\sg\ta^\prime}{2C}
+ \frac{(d+1)A\ta}{4C} \, .
\eeq
By doing the same change of variable from $\sg$ to $z$ as before
(where $z$ is given in eq. (\ref{eq:zdef})), it can be transformed into 
a more recognisable form given by,
\beq
\label{eq:aldiff_MVz}
z(1-z) \frac{{\rm d}^2 \al}{{\rm d}z^2}
+ \biggl[
\frac{d}{2}+1 - (d+2)z
\biggr] \frac{{\rm d}\al}{{\rm d}z} - d\al
= \frac{d(d-1)}{R} \biggl[
\frac{z(1-z)}{2} \frac{{\rm d}\ta}{{\rm d}z}
+ \frac{d+1}{4} (1-2z) \ta
\biggr] \, .
\eeq
This is a second order in-homogenous differential equation 
of hyper-geometric form with the parameters,
\beq
\label{eq:alpara_MV}
a_1^\prime = d \, ,
\hspace{5mm}
b_1^\prime = 1 \, ,
\hspace{5mm}
c_1^\prime = \frac{d}{2}+1 \, .
\eeq
Being second order in nature, it will have two independent 
homogenous solution. However it will also have a particular solution.
In the following (for a while) we will work on the particular solution
and describe a procedure for obtaining one in the present case \cite{Hoker1999}.
The particular solution can be found by making a  
simple observation on a maximally symmetric background.
\beq
\label{eq:maxsym_propB}
\Box f = \frac{1}{r_d^2} \biggl[
z(1-z) \frac{{\rm d}^2 f}{{\rm d}z^2}
+ \left(\frac{d}{2} - dz \right) \frac{{\rm d}f}{{\rm d}z}
\biggr] \, ,
\eeq
where $r_d^2 = d(d-1)/R$ gives the radius of the deSitter space
in $d$-dimensions. It should be noted that it gives a very useful 
identity,
\beq
\label{eq:box_iden_m0_1}
\Box_d \frac{{\rm d}f}{{\rm d}z}
= \frac{r_{d-2}^2}{r_d^2} 
\frac{{\rm d}}{{\rm d}z} \left[
\Box_{d-2} + \frac{d-2}{r_{d-2}^2}
\right] f \, ,
\eeq
where $\Box_{d-2}$ is the laplacian in the $d-2$ dimensions. 
Similarly one can derive,
\beq
\label{eq:box_iden_m0_2}
\frac{{\rm d}}{{\rm d}z} \Box_d f
= \frac{r_{d+2}^2}{r_d^2} 
\left[\Box_{d+2} -\frac{d}{r_{d+2}^2} \right] \frac{{\rm d}f}{{\rm d}z} \, .
\eeq
These identities will help us in finding the particular solution for
the differential equation. It should be noted that by making use 
of these identities one can transform the eq. (\ref{eq:aldiff_MVz}) 
into the following form,
\beq
\label{eq:al_MV_partform}
r_{d+2}^2 \biggl[
\Box_{d+2} - \frac{d}{r_{d+2}^2}
\biggr] \al 
= \frac{r_d^2}{2} \biggl[
z(1-z) \frac{{\rm d}\ta}{{\rm d}z}
+ \frac{d+1}{2} (1-2z) \ta
\biggr] \, .
\eeq
Then using the identity in eq. (\ref{eq:box_iden_m0_2}) one 
can re-express the LHS in a simplified compact form and rewrite the 
differential equation for the variable $\al$ in the following manner,
\beq
\label{eq:al_MV_partform1}
\frac{{\rm d}}{{\rm d}z} \Box_d \smallint \al
= \frac{1}{2}\biggl[
z(1-z)\frac{{\rm d}\ta}{{\rm d}z}
+ \frac{d+1}{2} (1-2z) \ta
\biggr] \, ,
\eeq
where $\smallint \al = \smallint_0^z {\rm d}z^\prime \al(z^\prime)$. 
This will give a first order ODE for the variable $\al$.
This first order ODE is given by,
\beq
\label{eq:1stODEal_MV}
z(1-z) \frac{{\rm d}\al}{{\rm d}z}
+ \frac{d}{2} (1-2z) \al
=  \frac{r_d^2}{2} \int_0^z {\rm d}z^\prime
\biggl[
z(1-z) \frac{{\rm d}\ta}{{\rm d}z}
+ \frac{d+1}{2} (1-2z) \ta
\biggr]  \, .
\eeq
Now this can be very easily solved for the particular solution 
$\al(z)$ by method of quadrature.
The actual form of the particular solution is given by,
\beq
\label{eq:alpartMV_solnF}
\al_p(z) = \frac{r_d^2}{[z(z-1)]^{d/2}}
\int_0^z {\rm d}z^\prime \left[z^\prime(1-z^\prime)\right]^{d/2-1}
\int_0^{z^\prime} {\rm d}z^{\prime\prime}
\biggl[
z^{\prime\prime}(1-z^{\prime\prime}) \frac{{\rm d}\ta}{{\rm d}z^{\prime\prime}}
+ \frac{d+1}{2} (1-2z^{\prime\prime}) \ta
\biggr] \, .
\eeq
We performed the integration using {\it Mathematica}. 
The particular solution in four dimensions is found to be,
\bea
\label{eq:alpMVd4}
\al_p(z) &=& - \frac{R}{384\pi} \sec \left\{
\frac{\pi}{2} \sqrt{1-48\g}
\right\} \biggl[
\frac{24 \g z (z-1)+(a_1-2)z(2z-1)-1}{z^2(z-1)(a_1-2)} 
\notag \\
&&
\times \, {}_2F_1(a_1-2, b_1-2; c_1-2;z)
+ \frac{1-24\g z (z-1)}{z^2(z-1)(a_1-2)} \, {}_2F_1(a_1-2, b_1-1; c_1-2; z)
\notag \\
&&
- \frac{(6\g+1)(2z-3)}{(z-1)^2}
\biggr] \, ,
\eea
where $\g=m^2/R$. The full solution for the variable $\al$ 
consist of a part coming from homogenous equation also.
As the differential equation in (\ref{eq:aldiff_MVz}) is of second order,
therefore it will have two homogenous solution, however we will choose 
the one which has a singularity in the $z\to1$ limit and is regular 
for $z\to0$ (large distance limit). This condition allows us to write the 
homogenous solution as,
\beq
\label{eq:homo_al_MV}
\al_H(z) = p \times {}_2F_1(a_1^\prime, b_1^\prime; c_1^\prime;z) \,,
\eeq
where the parameters are written in the eq. (\ref{eq:alpara_MV}).
The full solution is then given by,
\beq
\label{eq:fullsolnform_al_MV}
\al(z) = \al_H(z) + \al_p(z) \, .
\eeq
In order to find the value of parameter $p$ in $\al_H(z)$, 
we require that the $z\to1$ limit
(the short distance behaviour) of the full $\al(z)$ should match with the
strength of singularity of the transverse flat space-time propagator.
In flat space-time the leading contribution proportional to $g_\mnp$
in the transverse propagator is given by,
\beq
\label{eq:flatprop_MV}
G_\mnp (x, x^\prime) 
\sim
\frac{1}{2} \frac{1}{(4\pi)^{d/2}} \G\left(\frac{d}{2}-1\right)
\left(\frac{2}{\sg}\right)^{d/2-1} g_\mnp \, .
\eeq
It should be noticed that in the $z\to1$ limit, the leading 
singularity in the homogenous and particular solution,
namely $\al_H(z)$ and $\al_p(z)$ respectively, 
goes like $\sim(1-z)^{-d/2}$, however the leading singularity in the 
flat space-time transverse propagator goes like $\sim(1-z)^{1-d/2}$.
In order for the $z\to1$ limit of the full solution to 
match the singularity of flat space-time propagator, it is required that the 
leading singularity in both the homogenous and particular solution 
should cancel each other. This fetches the value for the parameter $p$ 
in four dimensions to be,
\beq
\label{eq:valP_MV}
p = -\frac{R}{4} \biggl[
\frac{1}{384\pi^2} \frac{1}{\g} - \frac{6\g+1}{32\pi} 
\sec \left\{
\frac{\pi}{2} \sqrt{1-48\g}
\right\}
\biggr] \, .
\eeq
It is noticed that once the leading singularity 
terms in the full solution for $\al$ are made zero, 
then the coefficient of next to leading singularity terms 
match with the strength of the singularity of the flat 
space-time massive propagator given in eq. (\ref{eq:flatprop_MV}). 
The function $\bt(z)$ can be determined using the eq. (\ref{eq:btinta_MV})
which when the transformation from $\sg$ to $z$ is made 
acquires the following form,
\beq
\label{eq:btsoln_MV}
\bt(z) = \frac{R}{2d(d-1) (\acos\sqrt{z})^2}
\biggl[
z(1-z) \frac{{\rm d} \al}{{\rm d}z} 
+ (1-z) \al - \frac{d(d-1)}{2R} z(1-z) \ta
\biggr] \, .
\eeq
In four dimensions the relevant functions are,
\bea
\label{eq:relnFunc_MV}
\ta(z) &=& \frac{R^2 \G(a_1) \G(b_1)}{2304 \pi^2} 
\, {}_2F_1(a_1, b_1; c_1; z) \, 
\notag \\
\tau(z) &=& -\frac{R^3\G(a_1) \G(b_1)}{27648 \pi^2} 
\frac{z-1}{(\acos\sqrt{z})^2} \biggl[
z(2 \g+1) \times _2F_1(a_1+1, b_1+1; c_1+1; z)
+ {}_2F_1(a_1, b_1; c_1, z)
\biggr] \notag \\
\al(z) &=& \frac{R}{4608\pi^2} \biggl[
\frac{12 \pi \sec \left(\pi(a_1-b_1)/2\right)}{z^2(z-1)(a_1-b_1+1)} \bigl\{
(-48\g z(z-1) -(a_1-b_1+1)z(2z-1) +2) 
\notag \\
&&
\times \, {}_2F_1(a_1-2, b_1-2; c_1-2;z)
+ (48 \g z(z-1) -2)\times {}_2F_1(a_1-2, b_1-1; c_1-2;z) \bigr\}
\notag \\
&&
+ \frac{2z-3}{\g(z-1)^2}
\biggr] \, .
\eea
Here for simplicity we have not written the expression for 
the function $\bt(z)$. It should be noticed that all the functions 
$\ta(z)$, $\tau(z)$, $\al(z)$ and $\bt(z)$ correctly 
reproduce the short distance limit, which matches with the 
strength of singularity in flat space-time. In the limit when 
the separation between the points is antipodal ($z\to0$ limit), 
the functions are regular. However 
they do not go to zero in the $z\to0$ limit. The limiting value 
of the massive transverse propagator at large distances is given by,
\beq
\label{eq:MVprop_z0}
\left. G^T_\mnp(x, x^\prime) \right|_{z\to0}
\sim \frac{R}{1536 \pi^2 \g} 
\biggl[12\pi \g(1+6\g) 
\times \sec(\frac{\pi}{2} \sqrt{1-48\g})-1\biggr]
\biggl[
g_\mnp + \frac{R}{6\pi^2} \sg_\mu \sg_{\nu^\prime}
\biggr] \, .
\eeq
From this it will look like that the $\g\to0$ limit is singular. But this 
is misleading. When the $\g\to0$ limit is taken one gets a perfectly 
regular answer. This is given by,
\beq
\label{eq:m0limGmnT}
\left. G^T_\mnp(x, x^\prime) \right|_{z\to0, \g\to 0}
\sim
-\frac{R}{256 \pi^2} \biggl[
g_\mnp + \frac{R}{6\pi^2} \sg_\mu \sg_{\nu^\prime}
\biggr] \, .
\eeq
This matches with the $z\to0$ limit of the transverse propagator 
in the massless vector field case (covered in the next section). 
It should also be noticed that this correlation is proportional to 
Ricci-scalar $R$, thereby implying that it vanishes in the $R\to0$ limit. 
The massless limit and the $z\to0$ limit completely commute 
with each other in the transverse part of the green's function. 

\subsection{Longitudinal Part}
\label{LGMV}

We then consider the longitudinal part of the full green's function
for the massive vector field. This can be obtained using 
eq. (\ref{eq:TLreln_EQM_MV}) or (\ref{eq:supconst_MV}). 
We use the later in order to obtain the longitudinal part of the 
green's function. It should be indicated that the past 
studies conducted for the massive vectors \cite{AllenJacob1985,
Higuchi2014} takes into account the longitudinal part also.
While the former computes the full green's function without 
distinguishing the transverse and longitudinal part, the 
later computes the propagator via canonical methods. 
The investigation done in \cite{Tsamis2006} however 
doesn't take into account the longitudinal part. 
Here we compute the longitudinal part of the green's 
function explicitly using the imposition of the 
supplementary constraint given in eq. (\ref{eq:supconst_MV}).

In the eq. (\ref{eq:TLreln_EQM_MV}) we first note that the longitudinal
part appearing is completely dictated by the function $G(\sg)$ alone.
Double derivatives of this like $\nb_\mu \nb_{\nu^\prime} G(\sg)$,
can be easily computed on a maximally symmetric background as,
\beq
\label{eq:Gexp_longm0}
\nb_\mu \nb_{\nu^\prime} G 
= C G^\prime g_\mnp + \biggl(
G^{\prime\prime} g_\mnp + \frac{1+C}{2\sg} G^\prime
\biggr) \sg_\mu \sg_{\nu^\prime} 
= \al_L g_\mnp + \bt_L \sg_\mu \sg_{\nu^\prime} \, ,
\eeq
where we have used the expansion of $\sg_\mnp$ on 
maximally symmetric space and wrote it in terms of $g_\mnp$ and
$\sg_\mu\sg_{\nu^\prime}$ using eq. (\ref{eq:sgmnp_C})
and in the second equality we just introduced new variables 
$\al_L$ and $\bt_L$ as the coefficients of $g_\mnp$ and
$\sg_\mu \sg_{\nu^\prime}$ respectively. It should be noticed that
due the dependence of $\al_L$ and $\bt_L$ on $G$, one can write
$\bt_L$ as follows,
\beq
\label{eq:btL_inalL_m0}
\bt_L = C^{-1} \left(\al_L^\prime
+ \frac{A+C}{2\sg} \al_L \right) \, .
\eeq
The longitudinal part of the green's function satisfies the 
following condition, which arises from the supplementary 
constraint given in eq. (\ref{eq:supconst_MV}),
\beq
\label{eq:LGMV_const}
\nb^\mu \nb_\mu \nb_{\nu^\prime} G(\sg) =0 \, .
\eeq
This translates in to a condition on the functions $\al_L$ and $\bt_L$
which is given by,
\beq
\label{eq:albt_LGMV}
\al_L^{\prime} - 2\sg \bt_L^\prime 
+ \frac{(d-1)(A+C)}{2\sg} \al_L - \left((d-1)A+2 \right) \bt_L =0 \, .
\eeq
Making use of the expression for $\bt_L$ from eq. (\ref{eq:btL_inalL_m0}),
we eliminate $\bt_L$ and its derivative from eq. (\ref{eq:albt_LGMV})
thereby obtaining an equation for the function $\al_L(\sg)$.
This is given by,
\beq
\label{eq:alL_MV_diff}
2 \sg \al_L^{\prime\prime} 
+ [(d+1) A+1] \al^\prime - \frac{R}{d-1} \al_L =0 \, .
\eeq
This should be compared with the corresponding equation (\ref{eq:alLeq_m0})
for the function $\al_L$ in the massless vector case. There 
the RHS is dependent on the gauge parameter $\lam$, and 
vanishes in the case of Landau gauge ($\lam=0$). 
This also implies that the present case is like 
solving the system in the Landau gauge, except the boundary 
condition imposed here will be different. Making transformation 
of variable from $\sg$ to $z$ in the eq. (\ref{eq:alL_MV_diff}), it
acquires the following form,
\beq
\label{eq:alL_MV_diffz}
z(1-z) \frac{{\rm d}^2\al_L}{{\rm d}z^2}
+ \biggl[\frac{d}{2} + 1 -z(d+2) \biggr] \frac{{\rm d}\al_L}{{\rm d}z}
- d \al_L = 0 \, .
\eeq
This is a second order homogenous ODE of hypergeometric form
with the parameters given in eq. (\ref{eq:alpara_MV}). This 
will again have two linearly independent solutions. However 
we will chose the one which is singular at $z\to1$ and regular 
at $z\to0$. This will imply
\beq
\label{eq:alL_MVsoln}
\al_L(z) = q_L \times {}_2F_1(a_1^\prime, b_1^\prime; c_1^\prime; z) \, ,
\eeq
where $q_L$ is the parameter to be fixed by comparing the short
distance behaviour of $\al_L(z)$ with the singularity structure of the
longitudinal part of the massive green's function in flat space-time 
background given in eq. (\ref{eq:GmnpLMV}). This gives the
value of $q_L$ to be,
\beq
\label{eq:qL_MV}
q_L = - \frac{1}{m^2} \frac{\G(d)}{\G(d/2)} \frac{(4\pi)^{-d/2}}{d} 
\biggl[
\frac{R}{d(d-1)}
\biggr]^{d/2} \, . 
\eeq
The factor of $1/m^2$ in front of the expression for $q_L$ causes problems
when the massless limit is taken. While the massless limit of the 
transverse propagator is completely well defined, the same is 
not true for the longitudinal part of the green's function. However,
this shouldn't come as a surprise, as in flat space-time also the 
longitudinal part has the same problem. This is due to fact that
when $m^2\to0$ at the level of path-integral, then the resulting 
action is gauge-invariant, and need to be gauge fixed in order to
have a well-defined path-integral. This is one source of the 
problem. The other source of the problem is the presence of 
zero modes. The green's function when computed while 
summing over the various eigen-modes, it is noticed that
mode corresponding to zero eigenvalue will cause problems
and will lead to divergences in the propagator. This is 
the famous infrared problem in the deSitter background. 
The full longitudinal part of the green's function for the massive 
vector field in four dimensions is given by,
\bea
\label{eq:albt_L_MVd4}
\al_L(z) &=& \frac{R}{4608 \pi^2} \frac{1}{\g} \frac{2z-3}{(1-z)^2} \, ,
\\
\bt_L(z) &=& \frac{R^2}{110592 \pi^2} \frac{1}{\g} \frac{z-3}{(1-z)^2 
(\acos\sqrt{z})^2 } \, .
\eea
The function $\al_L(z)$ can be integrated in order to work out the expression 
for $G(z)$ introduced in eq .(\ref{eq:GTV_decomp}). This can be done 
by integrating the following relation between $\al_L(z)$ and $G(z)$:
\beq
\label{eq:Geq_alL_m0}
\frac{{\rm d}G(z)}{{\rm d}z} = \frac{2d(d-1)}{R} \al_L \, .
\eeq
In four dimensions when this is integrated 
we get the following $G(z)$ (modulo integration constant)
\beq
\label{eq:Gz_MV}
G(z) = \frac{1}{192 \pi^2} \frac{1}{\g} \biggl[
-\frac{1}{1-z} + 2 \ln(1-z)
\biggr] \, .
\eeq
This should be compared with the analogous function obtained 
for massless case in eq. (\ref{eq:G_soln_4Dm0}). $G(z)$ being 
proportional to $1/\g$ reminds us of the ill-defined massless 
limit of the longitudinal part of the massive green's function. 
This is a source of trouble. The $z\to0$ limit of the 
longitudinal part is easy to compute. This when 
combined with the $z\to0$ limit of the transverse part of the 
massive propagator, the $z\to0$ limit of the full propagator is 
given by,
\beq
\label{eq:Gmnp_MV_z0}
\left. G_\mnp(z) \right|_{z\to0} \sim 
- \frac{R}{256\pi^2} \biggl[
1 + \frac{1}{6\g}
\biggr] \biggl[
g_\mnp + \frac{R}{6\pi^2} \sg_\mu \sg_{\nu^\prime} 
\biggr] \, .
\eeq
Here the term proportional to the $1/\g$ is coming from the 
longitudinal part. It should be mentioned that the 
massless limit and $z\to0$ limit commute in the case of transverse 
part of the massive green's function and is computable,
while the same is not true for longitudinal part. This will also
imply that massless limit of the full propagator cannot be 
taken, where the source of the problem is arising from the 
longitudinal part of the massive green's function. 
The other interesting point to note is the negative sign 
in front of the correlation between the fields at antipodal 
separation. This hints that perhaps there are some edge 
states whose contributions need to be taken correctly.

\section{Massless Vector Field}
\label{massless_sp1}

Now we consider the case for massless vector fields. 
Massless vector fields occur in many context. For example the 
photon field is described by a massless vector, and the same is true
of the massless gluon field. For simplicity we consider the case of
massless photon field, as the gauge group is simpler for them.
The euclidean action for the photon field is given by, 
\beq
\label{eq:act_photon_c}
S = \int {\rm d}^dx \sqrt{g} \biggl[
\frac{1}{4} F_\mn F^\mn + \frac{1}{2\lam} (\nb_\mu A^\mu)^2
\biggr] \, ,
\eeq
where $F_\mn = \nb_\mu A_\nu - \nb_\nu A_\mu$ 
as defined before is the field strength 
tensor for the vector field $A_\mu$. The action for the gauge field is 
given by the first term of eq. (\ref{eq:act_photon_c}), which has 
a $U(1)$ gauge invariance. When the path-integral is written for the 
$U(1)$ invariant action, it is noticed that the measure over the field 
is not well defined and needs to be gauge fixed in order to prevent the 
over counting of gauge orbits. For this reason we have added the 
second term in the eq. (\ref{eq:act_photon_c}) as the gauge fixing action.
This is done via the Faddeev-Popov procedure and will generate ghost
determinants in the analysis. However these determinants will 
not effect our study of the green's function, as they don't play any role
in their analysis. In the eq. (\ref{eq:act_photon_c}) $\lam$ is the gauge 
parameter. One can then follow the steps of the previous section
and notice that the equation satisfied by the two-point function is given by
\beq
\label{eq:eqmotion_U1}
W^\mn \langle A_\nu A_{\nu^\prime} \rangle 
= \biggl[
\left(- \Box + \frac{R}{d} \right) g^\mn 
+\left(1-\frac{1}{\lam} \right) \nb^\mu \nb^\nu
\biggr]  \langle A_\nu A_{\nu^\prime} \rangle = 
\frac{g^\mu{}_{\nu^\prime} \de(x-x^\prime)}{\sqrt{g^\prime}} \, ,
\eeq
where $\langle A_\nu A_{\nu^\prime} \rangle = G_\mnp(x, x^\prime)$ is the
green's function for the massless propagator. Again as in previous
section we solve for the green's function for the non-coincident 
points and then match with the flat space-time case 
at short distances, which is like employing the correct 
boundary condition. From eq. (\ref{eq:eqmotion_U1}) one can 
determine $\nb^\mu \langle F_\mn A_{\nu^\prime} \rangle$ 
analogous to eq. (\ref{eq:nabF}), which is given by,
\beq
\label{eq:Fdiv_m0}
\nb^\mu \langle F_\mn A_{\nu^\prime} \rangle 
= -\frac{1}{\lam} \nb^\mu \nb^\nu \langle A_\nu A_{\nu^\prime} \rangle \, .
\eeq
These two equations will help us in determining the green's function
for the massless vector fields on a DeSitter background. 
From now onwards we will follow some of the steps that were 
used in the computation of propagator of massive vector fields.
As in eq. (\ref{eq:GTV_decomp}) we decompose the green's function 
$G_\mnp$ for massless vector field in to transverse and
longitudinal parts namely $G^T_\mnp$ and
$\nb_\mu \nb_{\nu^\prime} G$ respectively, where the transverse
part satisfies eq. (\ref{eq:trans_G}). Plugging this decomposition
in the eq. (\ref{eq:eqmotion_U1}), gives us relation between the 
longitudinal and transverse part of the massless green's function to be,
\beq
\label{eq:longG_m0}
\nb_\mu \nb_{\nu^\prime} \Box G
= \lam \biggl(-\Box + \frac{R}{d} \biggr) G^T_\mnp \, .
\eeq
From this we notice that the longitudinal part will be proportional to 
the gauge parameter $\lam$. Therefore in Landau gauge it will be 
zero. This equation can be rewritten in the following form,
\beq
\label{eq:longG_m0_form1}
\biggl(-\Box + \frac{R}{d} \biggr)\biggl[
\nb_\mu \nb_{\nu^\prime} G + \lam G^T_\mnp
\biggr] =0 \, .
\eeq
This implies as before that solving for one, will give us the other. 

\subsection{Transverse Part}
\label{eq:TRGm0}

We decide to solve for the transverse part of the green's function 
first. On a maximally symmetric background the transverse part
can be written as in eq. (\ref{eq:TV_GMV}). We then 
consider the quantity $\langle F_\mn F^{\mu^\prime\nu^\prime}\rangle$.
As this quantity is gauge invariant therefore the longitudinal 
part of $G_\mnp$ drops out from this as before and it 
depends only on $G^T_\mnp$.
Then proceeding as before we write the quantity 
$\langle F_\mn F^{\mu^\prime\nu^\prime}\rangle$ as a linear 
combination of $g_{[\mu}{}^{[\mu^\prime}g_{\nu]}{}^{\nu\prime]}$
and $\sg_{[\mu}\sg^{[\mu^\prime}g_{\nu]}{}^{\nu\prime]}$ as in 
eq. (\ref{eq:FF_exp}), with the coefficients $\ta$ and $\tau$
as functions of $\sg$ respectively. Then $\ta$ and $\tau$ 
are given in terms of $\al$ and $\bt$ as in eq. (\ref{eq:albt_MV}). 
Considering the covariant divergence of 
$\langle F_\mn F^{\mu^\prime\nu^\prime}\rangle$
and making use of eq. (\ref{eq:Fdiv_m0}), it is found that
\beq
\label{eq:divFF_m0}
\nb^\mu \langle F_\mn F^{\mu^\prime \nu^\prime} \rangle
= -\frac{1}{\lam} \nb_\nu \Box \langle a F^{\mu^\prime\nu^\prime}
\rangle \, ,
\eeq
where we have decomposed the vector field $A_\mu$ in to 
transverse and longitudinal part as $A_\mu = A^T_\mu
+ \nb_\mu a$, with condition $\nb^\mu A^T_\mu=0$.
As we are on a maximally symmetric background, therefore 
the expectation value on the RHS can be written as a linear combination 
of $g^{\mu^\prime\nu^\prime}$ and $\sg^{\mu^\prime} \sg^{\nu^\prime}$.
However these quantities are symmetric in the pair $(\mu^\prime, \nu^\prime)$
thereby implying the quantity $\langle a F^{\mu^\prime\nu^\prime}
\rangle$ will be zero. This give us a supplementary condition,
\beq
\label{eq:supcond_m0}
\nb^\mu \langle F_\mn F^{\mu^\prime \nu^\prime} \rangle=0 \, .
\eeq
This should be contrasted with the similar condition that is 
obtained in the massive case (check eq. (\ref{eq:covDiv_FMV})).
It should be emphasised that the two equations (\ref{eq:covDiv_FMV}
and \ref{eq:supcond_m0}) agree with each other in the massless limit.
The condition stated in eq. (\ref{eq:supcond_m0}) will translate 
in to a condition on the variables $\ta$ and $\tau$. 
By making use of eq. (\ref{eq:LHS_divFF}) it is noticed that
eq. (\ref{eq:supcond_m0}) will imply
\beq
\label{eq:supcond_m01}
- \ta^\prime + \sg \tau^\prime
- \frac{(d-2)(A+C)}{2\sg} \ta
+ \frac{(d-2)A+2}{2} \tau =0 \, ,
\eeq
where the `prime' here denotes the derivative of the variable 
with respect to the argument $\sg$. From the supplementary 
condition stated in eq. (\ref{eq:supcond_m01}) one can 
eliminate $\tau$ by making use of the expressions given in 
eq. (\ref{eq:albt_MV}). After doing the algebra and some 
simplifications one acquires a simple looking equation for the 
variable $\ta$. This is given by,
\beq
\label{eq:ta_m0}
2\sg \ta^{\prime\prime}
+ [(d+1)A+1] \ta^\prime - \frac{2R}{d} \ta =0 \, .
\eeq
This should be compared with the equation for the function $\ta(\sg)$
in the case of massive vector field, given in eq. (\ref{eq:taeq_FFMV}).
The two equations are exactly the same except for the presence of 
mass term in the massive vector case.
Using the expression for $A$ on a deSitter space from eq. (\ref{eq:ACvalds}) 
and making the change of variable from $\sg$ to $z$ by making use 
of the definition for $z$ given in eq. (\ref{eq:zdef}), 
we will get a differential equation for the variable $\ta$ 
in terms of $z$. This equation will have a hyper-geometric form 
and is given by following,
\beq
\label{eq:ta_HG_m0}
z(1-z) \frac{{\rm d}^2 \ta}{{\rm d}z^2}
+ \biggl[\frac{d}{2} +1 - (d+2)z \biggr] 
\frac{{\rm d}\ta}{{\rm d}z} - 2(d-1) \ta =0 \, .
\eeq
One should compare this equation with the equation for
$\ta(z)$ in eq. (\ref{eq:ta_z_FFMV}). The two are same except for 
mass term in later. This differential equation is of Hyper-geometric 
form with the parameters,
\beq
\label{eq:tapara_GTm0}
a_1 = d-1 \, , \hspace{5mm}
b_1 = 2 \, , \hspace{5mm}
c_1 = \frac{d}{2}+1 \, .
\eeq
The differential equation in (\ref{eq:ta_HG_m0}) being second order, 
will have two linearly independent solutions. But its only in the Bunch-Davis 
vacuum that we get the propagator with the required physical 
properties {\it i.e} 1) having the same short distance singularity as
in flat space-time, 2) regularity at large distances. These two conditions 
helps us in finding the right solution for the homogenous 
equation. This can be written as,
\beq
\label{eq:ta_soln_m0}
\ta(z) = q \times _2F_1 (a_1, b_1; c_1; z) \, ,
\eeq
where $q$ is determined by comparing the value of 
$\ta(z)$ in the limit $z\to1$ with the flat space-time 
expression for $\ta(z)$ given in appendix \ref{masslessvecflat}
in eq. (\ref{eq:FF_flatm01}). The value of $q$ is,
\beq
\label{eq:q_ta_m0}
q = \frac{2}{(4\pi)^{d/2}} \biggl[
\frac{R}{d(d-1)}
\biggr]^{d/2} \frac{\G(d-1)}{\G(d/2+1)} \, .
\eeq
On comparing this solution for the function $\ta(z)$ with the 
corresponding one in the massive vector case eq. (\ref{eq:greenFFMV}), 
it is noticed that the later matches the former if the mass $m^2\to0$ in the later. 
Once $\ta(z)$ has been found, it can be used to compute the value 
of $\tau(z)$ using the eq. (\ref{eq:tauZeq}). This is given by,
\bea
\label{eq:tauZm0}
\tau(z) &=& \frac{2\G(d-1)}{(4\pi)^{d/2}} \left(\frac{R}{d(d-1)}\right)^2
\frac{z-1}{(\acos\sqrt{z})^2} 
\biggl[
\frac{{}_2F_1(d-1,2;d/2+1;z)}{\G(d/2+1)}
\notag \\
&&
+ \frac{(d-1)z{}_2F_1(d,3;d/2+2;z)}{\G(d/2+2)}
\biggr] \, .
\eea
This also matches with the function $\tau(z)$ found in the massive 
vector case in the mass $m^2\to0$ limit. 
Once we find the function $\ta(z)$, it can be used to compute the 
functions $\al$ and $\bt$. We make use of the transversality 
constraint stated in eq. (\ref{eq:trans_G}), to write a 
relation between $\al$ and $\bt$. This is given in eq. (\ref{eq:albt_relt_MV}).
Using this relation and by eliminating $\bt$ and $\bt^\prime$
from the equation by using eq. (\ref{eq:btinta_MV}), we arrive 
at an equation for the function $\al$. This is given by,
\beq
\label{eq:aleq_m0_GT}
2\sg \al^{\prime\prime}
+ [(d+1)A+1] \al^\prime - \frac{R}{d-1} \al
= \frac{\sg}{2C} \ta^\prime
+ \frac{(d+1)A}{4C} \ta \, .
\eeq
This should be compared with the corresponding equation
(\ref{eq:aldiff_MV}) for the function $\al$ in the case of massive 
vector. It is worth noting that the LHS of the two equations 
are exactly the same. However although the RHS has the same 
structure but they are different as in massive case, the function 
$\ta$ has mass dependence. But in the limit of $m^2\to0$ the
two equation are identical. We can do a change of variable from $\sg$
to $z$ as previously done and obtain a differential equation for $\al(z)$.
The equation for $\al(z)$ is given by,
\beq
\label{eq:alGT_m0z}
z(1-z) \frac{{\rm d}^2 \al}{{\rm d}z^2}
+ \biggl[
\frac{d}{2}+1 - (d+2)z
\biggr] \frac{{\rm d}\al}{{\rm d}z} - d\al
= \frac{d(d-1)}{R} \biggl[
\frac{z(1-z)}{2} \frac{{\rm d}\ta}{{\rm d}z}
+ \frac{d+1}{4} (1-2z) \ta
\biggr] \, .
\eeq
This is an in-homogenous differential equation of hyper-geometric form
with the parameters,
\beq
\label{eq:alpara_m0}
a_1^\prime = d \, ,
\hspace{5mm}
b_1^\prime = 1 \, ,
\hspace{5mm}
c_1^\prime = \frac{d}{2}+1 \, .
\eeq
This equation being a second order in-homogenous differential 
equation, will have two linearly independent homogenous solution
beside a particular solution. The particular solution is found by 
using the same steps as in previous section, namely the 
eq. (\ref{eq:maxsym_propB} - \ref{eq:1stODEal_MV}), and write the 
particular solution as in eq. (\ref{eq:alpartMV_solnF}) to be,
\beq
\label{eq:alpart_GT_m0}
\al_p(z) = \frac{r_d^2}{[z(z-1)]^{d/2}}
\int_0^z {\rm d}z^\prime \left[z^\prime(1-z^\prime)\right]^{d/2-1}
\int_0^{z^\prime} {\rm d}z^{\prime\prime}
\biggl[
z^{\prime\prime}(1-z^{\prime\prime}) \frac{{\rm d}\ta}{{\rm d}z^{\prime\prime}}
+ \frac{d+1}{2} (1-2z^{\prime\prime}) \ta
\biggr] \, ,
\eeq
where $\ta(z)$ is given in eq. (\ref{eq:ta_soln_m0}). In four dimensions
the particular solution is given by,
\beq
\label{eq:alpart_4d_m0}
\al_p(z) = \frac{R}{768\pi^2} \biggl[
-\frac{2z^2-z-2}{z(1-z)^2} 
+ \frac{2(2z+1) \ln(1-z)}{z^2}
\biggr] \, .
\eeq
The full solution for the function $\al(z)$ also contains a
homogenous part. This homogenous part is given by
\beq
\label{eq:alhomo_m0}
\al_H(z) = p \times {}_2F_1(a_1^\prime, b_1^\prime; c_1^\prime;z) \, ,
\eeq
where the parameters are written in eq. (\ref{eq:alpara_m0}). 
The value of the parameter $p$ is determined as before, by requiring 
that the strength of singularity of the full solution should match the 
strength of singularity of the transverse propagator in flat space-time. 
This requirement implies that the leading singularity in the 
particular and the homogenous solution, which goes like
$\sim(1-z)^{-d/2}$ should cancel each other. This imposition
gives the value of $p$ in four dimensions to be,
\beq
\label{eq:p_m0_al}
p = -\frac{R}{256\pi^2} \, .
\eeq
Once the value of $p$ is fixed, it is noticed that the singularity 
proportional to $(1-z)^{1-d/2}$ in the full $\al(z)$, matches the
term proportional to $g_\mnp$ in the transverse contribution of the flat 
space-time propagator, where the later is given by,
\beq
\label{eq:flat_al_m0}
G^{\rm flat}_\mnp \sim
\frac{1}{d-2} \frac{1}{(4\pi)^{d/2}} \left(\frac{\sg}{2} \right)^{1-d/2}
g_\mnp \, .
\eeq
The solution to the function $\bt(z)$ can be found using 
eq. (\ref{eq:btsoln_MV}). In four dimensions the functions 
$\al$ and $\bt$ are therefore given by,
\bea
\label{eq:albtm0_d4}
\al(z) &=& \frac{R}{384\pi^2}\biggl[
\frac{1}{z(1-z)} + \frac{(2z+1) \ln(1-z)}{z^2}
\biggr] \, ,
\notag \\
\bt(z) &=& -\frac{R^2}{9216\pi^2 (\acos\sqrt{z})^2}
\biggl[
\frac{1}{z(1-z)} + \frac{(1-z) \ln(1-z)}{z^2}
\biggr] \, .
\eea
Once $\al(z)$ and $\bt(z)$ are found, we have the information about
the transverse part of the Green's function $G^T_\mnp$. This can 
then be used to determine the longitudinal part $G(z)$ of the green's function.

\subsection{Longitudinal Part}
\label{eq:LGGm0}

The longitudinal part of the green's function can 
be obtained from eq. (\ref{eq:longG_m0_form1}). 
Solving for the longitudinal part $G$ is a bit involved.
Writing the longitudinal part $\nb_\mu\nb_{\nu^\prime}G$ as in 
eq. (\ref{eq:Gexp_longm0}), we note the relation between 
the longitudinal functions $\al_L$ and $\bt_L$ given in
eq. (\ref{eq:btL_inalL_m0}). As $G^T_\mnp$ has been written in terms of $\al$ and $\bt$ 
in eq. (\ref{eq:TV_GMV}), therefore we compute the action of
the operator $(\Box - R/d)$ on $G^T_\mnp$. In particular we consider the 
coefficient of $g_\mnp$. This coefficient can then be
simplified using eq. (\ref{eq:albt_MV} and \ref{eq:aldiff_MV}). 
Finally after a bit of manipulation this is given by,
\beq
\label{eq:Opact_Gmnp}
\left. \left(\Box - \frac{R}{d} \right) G^T_\mnp \right|_{g_\mnp}
= \frac{\sg}{2C} \ta^\prime + \frac{(d-1)A}{4C}  \ta \, .
\eeq
Then we consider the coefficient of $g_\mnp$ in the 
$(\Box-R/d) \nb_\mu \nb_{\nu^\prime} G$. As $\nb_\mu \nb_{\nu^\prime} G$
has an expansion given in eq. (\ref{eq:Gexp_longm0}), therefore it
is found that coefficient proportional to $g_\mnp$ is given by following, 
\beq
\label{eq:gmnp_coeff_L}
\left. \left(\Box -\frac{R}{d} \right) \nb_\mu \nb_{\nu^\prime} G \right|_{g_\mnp}
= 2\sg \al_L^{\prime\prime} 
+ ((d-1)A+1) \al_L^\prime 
- \frac{(A+C)^2}{2\sg} \al_L - \frac{R}{d} \al_L
+ 2 \bt_L AC \, .
\eeq
Expressing $\bt_L$ in terms of $\al_L$ using eq. (\ref{eq:btL_inalL_m0})
and using eq. (\ref{eq:longG_m0_form1}), we obtain an equation
for $\al_L$ to be,
\beq
\label{eq:alLeq_m0}
2\sg \al_L^{\prime\prime}
+ [(d+1)A+1] \al_L^\prime - \frac{R}{d-1} \al_L
= - \lam \biggl[
\frac{\sg}{2C} \ta^\prime + \frac{d-1}{4} \frac{A}{C} \ta
\biggr] \, .
\eeq
This is the differential equation for the longitudinal part
of the Green's function. This should be compared with the 
analogous equation obtained for $\al_L(\sg)$ in the massive vector 
case (see eq. (\ref{eq:alL_MV_diff})), where the RHS is 
zero, and can be obtained by putting $\lam=0$ (Landau 
gauge) in the eq. (\ref{eq:alLeq_m0}). 
This can be rewritten by a change of variable from 
$\sg$ to $z$ using eq. (\ref{eq:zdef}), after which it acquires the 
following form,
\beq
\label{eq:alLdiff_z_m0}
z(1-z) \frac{{\rm d}^2\al_L}{{\rm d}z^2}
+ \biggl[\frac{d}{2} + 1 -z(d+2) \biggr] \frac{{\rm d}\al_L}{{\rm d}z}
- d \al_L
= -\lam\frac{d(d-1)}{2R} \biggl[z(1-z) \frac{{\rm d}\ta}{{\rm d}z}
+(1-2z) \frac{d-1}{2} \ta
\biggr] \, .
\eeq
This is a second order in-homogenous differential equation of 
hyper-geometric form with the parameters same as in 
eq. (\ref{eq:alpara_m0}). This equation will again have a 
homogenous solution and a particular solution. The particular 
solution is obtained as before, and is given by 
\beq
\label{eq:part_alL_m0}
\al^p_L(z) = -\frac{\lam r_d^2}{[z(z-1)]^{d/2}}
\int_0^z {\rm d}z^\prime \left[z^\prime(1-z^\prime)\right]^{d/2-1}
\int_0^{z^\prime} {\rm d}z^{\prime\prime}
\biggl[
z^{\prime\prime}(1-z^{\prime\prime}) \frac{{\rm d}\ta}{{\rm d}z^{\prime\prime}}
+ \frac{d-1}{2} (1-2z^{\prime\prime}) \ta
\biggr] \, ,
\eeq
while the homogenous solution is given by
\beq
\label{eq:homo_alL_form}
\al_L^H(z) = p_L \times {}_2F_1(a_1^\prime, b_1^\prime; c_1^\prime;z) \, ,
\eeq
where the parameter $p_L$ is to be fixed later. In four dimension one 
can solve for the particular solution by using the known form of the 
function $\ta(z)$ given in eq. (\ref{eq:ta_soln_m0}). This is given by
\beq
\label{eq:alLpart_m0}
\al^p_L(z) = -\frac{\lam R}{768\pi^2}\biggl[
-\frac{2z^2-z-2}{z(1-z)^2} + \frac{2(2z+1)\ln(1-z)}{z^2}
\biggr] \, .
\eeq
The parameter $p_L$ in the homogenous solution is determined 
by requiring that the singularity proportional to $(1-z)^{-d/2}$
in the homogenous solution should cancel the corresponding 
singularity in the particular solution. This fixes the value for 
$p_L$ in four dimensions to be,
\beq
\label{eq:longcoeff_alL}
p_L = \frac{11\lam R}{2304\pi^2} \, .
\eeq
Once this is fixed it is noticed that the next to leading singularity,
which is proportional to $(1-z)^{1-d/2}$ in the full solution for the 
$\al_L$ matches exactly the longitudinal contribution in the massless 
propagator in flat space-time. In four dimensions the full solution 
to the longitudinal $\al_L$ (and the solution to $\bt_L$ 
using eq. (\ref{eq:btL_inalL_m0})) is given by
\bea
\label{eq:albtLd4_m0}
\al_L(z) &=& \frac{R\lam}{1152\pi^2} 
\biggl[
\frac{4z-1}{z(1-z)} - \frac{(2z+1)\ln (1-z)}{z^2}
\biggr] \, ,
\notag \\
\bt_L(z) &=& \frac{R^2 \lam}{27648 \pi^2 (\acos\sqrt{z})^2}
\biggl[
\frac{2z^2-4z-1}{z(z-1)} + \frac{(1-z)\ln(1-z)}{z^2}
\biggr] \, .
\eea
Once knowledge of $\al_L(z)$ is acquired, it can be used to 
determine the function $G$ using eq. (\ref{eq:Geq_alL_m0}).
This when integrated in four dimensions using mathematica, gives 
the following function $G(z)$,
\beq
\label{eq:G_soln_4Dm0}
G(z) = \frac{\lam}{32\pi^2} \biggl[
\text{Li}_2(z) + \frac{(1-4z)\ln(1-z)}{2z}
\biggr] \, ,
\eeq
where $\text{Li}_2(z)$ is the poly-log function. 

In four dimensions the transverse part of the photon green's 
function is given by eq. (\ref{eq:albtm0_d4}) and the 
longitudinal part is given in eq. (\ref{eq:albtLd4_m0}). In either
case we notice that the $z\to1$ limit correctly matches the 
strength of singularity in flat space-time, while these functions 
are regular in the $z\to0$ limit. However in the $z\to0$ limit, these
function don't approach zero, instead goes to some constant. 
In the massless case the propagator in this limit (when the 
points are antipodal) has the following form,
\beq
\label{eq:AAprop_m0z0}
\left. G_\mnp (z) \right|_{z\to0}
\sim \biggl(-\frac{R}{256\pi^2} 
+ \frac{11R\lam}{2304\pi^2}
\biggr)
\biggl[
g_\mnp + \frac{R}{6\pi^2} \sg_\mu \sg_{\nu^\prime}
\biggr] \, .
\eeq
This should be compared with the massive vector case.
In the Landau gauge ($\lam=0$), the antipodal limit 
in the massless case matches exactly the $z\to0$ limit 
in the massive vector case, when the mass is taken to zero.
This holds as a consistency check in our computation.
It should be specified here that the antipodal point 
separation limit is not the true infrared limit which is 
usually witnessed in flat space-time. The reason being 
the euclidean deSitter is compact (basically $d$-sphere
in $d$-dimensions), therefore the maximum separation that
two points can have is half the circumference of circle 
joining the two points, not infinity which would be the case 
in Lorentzian space-times. 
However, the non-vanishing of this limit in our present case hints 
that there might be edge states living on the boundary, whose
effects need to be correctly incorporated. 

\section{Comparison with Past work}
\label{CPTW}

Here in this section we compare our results with the green's function 
that has been obtained in the past \cite{AllenJacob1985, Higuchi2014}.
We do this comparison for both the massive and massless vector fields.
It has been shown in \cite{Higuchi2014} that their results agree with 
the ones given in \cite{AllenJacob1985} when their Stueckelberg 
parameter goes to infinity in the case of massive vectors. As this 
agreement has already been established, therefore we only investigate the comparison 
of our results with the ones given in \cite{AllenJacob1985}. 
This is easily done, as the notation used by us is somewhat similar to the
ones used in \cite{AllenJacob1985}. The results of \cite{Tsamis2006} 
are seen to agree with the result of \cite{Higuchi2014} in the 
limit when the Stueckelberg parameter goes to zero.
The Stueckelberg parameter of \cite{Higuchi2014} play the 
role of gauge-fixing parameter in the gauge theories.
While the limit of this parameter going to infinity, it imposes 
the gauge fixing condition like a delta-function in the space of
fields, the limit of the parameter going to zero, doesn't 
impose any constraint on the vector fields at all. With 
this knowledge it is realised from the beginning that our results
in the massive vector case should agree with the ones 
in \cite{AllenJacob1985, Higuchi2014} in the limit of
Stueckelberg parameter going to infinity, while in the case 
of \cite{Tsamis2006}, the agreement should be only 
for the transverse part of the green's function
(as anyway the authors there are only computing the transverse 
green's function). 
As the results from \cite{Higuchi2014} agree with the ones in 
\cite{Tsamis2006} when the Stueckelberg parameter goes to zero,
this implies that the two agree on the transverse green's function.
But it has already been shown in \cite{Higuchi2014} that their results 
agree with \cite{AllenJacob1985} when the Stueckelberg
parameter goes to infinity, thereby implying that the two agree
not only on the transverse part, but also the longitudinal part.
So, all we will do in our paper is to check whether we 
agree with the results of \cite{AllenJacob1985} or not.

Instead of using the geodetic interval $\sg(x, x^\prime)$, the 
authors in \cite{AllenJacob1985} use the arc-length 
$\mu(x, x^\prime)$ as the bi-scalar. But $\sg(x, x^\prime)$ 
and $\mu(x, x^\prime)$ are related to each other by,
\beq
\label{eq:sgmu_rel}
\sg(x, x^\prime) = \frac{1}{2} \mu(x, x^\prime)^2 \, .
\eeq
Then the derivatives of $\sg$ and $\mu$ are related to each 
other by,
\beq
\label{eq:sgmu_derrel}
\sg_\al = \mu \, \mu_\al \, , 
\hspace{2mm}
\sg_\al \sg_\bt = \mu^2 \mu_\al \mu_\bt \, ,
\hspace{2mm}
\sg_\al \sg_{\bt^\prime} = \mu^2 \mu_\al \mu_{\bt^\prime} \, .
\eeq
Once this is specified, it is easy to translate the green's function
from one language to another. We start by computing 
$\al(\mu)$ and $\bt(\mu)$ by using the expressions written in 
eq. (3.18) of \cite{AllenJacob1985}. It should be noted that 
the bi-scalar $z(x, x^\prime)$ is same both the language. 
So, once the $\al(\mu)$ and $\bt(\mu)$ are computed, they 
are then written as function of $z(x, x^\prime)$.
The function $\al(z)$ and $\bt(z)$ are given by following,
\bea
\label{eq:albt_allen}
\al(z) &=& -\frac{(4\pi)^{-d/2}}{d \g} \frac{\G(a_1) \G(b_1)}{\G(d/2+2)}
\biggl(
\frac{R}{d(d-1)}
\biggr)^{d/2-1}
\biggl[
(d\g+2) z(1-z) {}_2F_1 (a_1+1, b_1+1; c_1+1;z)
\notag \\
&&
- \frac{d+2}{4} (2z-1) {}_2F_1(a_1, b_1; c_1; z)
\biggr]  \, ,
\notag \\
\bt(z) &=& -\frac{(4\pi)^{-d/2}}{d \g} \frac{\G(a_1) \G(b_1)}{\G(d/2+2)} 
\biggl(
\frac{R}{d(d-1)}
\biggr)^{d/2-1}
(1-z) \biggl[
(d\g+2) z {}_2F_1 (a_1+1, b_1+1; c_1+1;z)
\notag \\
&&
+ \frac{d+2}{2} {}_2F_1(a_1, b_1; c_1; z)
\biggr] \, .
\eea
Having written $\al(z)$ and $\bt(z)$ for massive vectors of 
\cite{AllenJacob1985} in arbitrary dimensions, we now extract 
information from it to seek comparison with our own results.
The sign ambiguity with them just indicates the different choice of conventions. 
From this we first extract the longitudinal piece. As the longitudinal 
piece is divergent as $m^2 \to0$, therefore it is easy to single out,
by expanding the above expressions in powers of $m^2$. The term 
proportional to $1/m^2$ is the longitudinal piece. 
This is given by,
\bea
\label{eq:allenLMV}
\al_L^{\rm Al}(z) &=& - \frac{2(4\pi)^{-d/2}}{R \g} \frac{\G(d)}{\G(d/2+2)}
\biggl(\frac{R}{d(d-1)}\biggr)^{d/2} \biggl[
z(1-z) {}_2F_1(a_1^\prime,b_1^\prime+2;c_1^\prime;z)
\notag \\
&&
+ \frac{1}{4} \left(\frac{d}{2} +1 \right) (1-2z) 
{}_2F_1(a_1^\prime-1, b_1^\prime+1;c_1^\prime;z)
\biggr] \, ,
\eea
where the superscript `Al' implies results obtained using the 
expression from \cite{AllenJacob1985}. 
This longitudinal piece matches with our longitudinal part of green's function 
mentioned in eq. (\ref{eq:albt_L_MVd4}) (after using 
some properties of the hyper-geometric functions). Now 
we make a comparison with the transverse part of the 
green's function. To do this, we go through a 
non-direct route {\it i.e} we consider the 
$\langle F_\mn F_{\mu^\prime \nu^\prime} \rangle$ 
correlation for the green's function given in 
\cite{AllenJacob1985}. The reason why we do this is
because $\langle F_\mn F_{\mu^\prime \nu^\prime} \rangle$ correlation 
only contains the transverse piece and not the longitudinal 
part. If we agree at this stage, then that implies complete 
agreement. To make this comparison, we first translate 
the $\al(\mu)$ and $\bt(\mu)$ of \cite{AllenJacob1985}
in to our notation. This is given by,
\beq
\label{eq:albt_allenTR}
\al(\sg) = \al^{\rm Al}(\sqrt{2\sg}) \, ,
\hspace{5mm}
\bt(\sg) = \frac{\bt^{\rm Al}(\sqrt{2\sg})}{2\sg} \, .
\eeq
Then we make use of the expression given in 
eq. (\ref{eq:albt_MV}) to compute $\ta(\sg)$ and 
$\tau(\sg)$. Making the change of variable 
from $\sg$ to $z$, allows us to simplify the algebra 
and make use of some of the properties of 
hypergeometric functions. Once this simplification is 
achieved via properties of hyper-geomertic functions 
then the resulting expression matches with our
result for $\ta(\sg)$. This will then imply matching 
of $\al(\sg)$ and $\bt(\sg)$ for the massive vector fields.

The case of massless vector fields is easy. We have here 
done the computation for the arbitrary gauge-fixing 
parameter $\lam$. We notice that in the gauge 
$\lam=1$ our results agree with the ones given in 
\cite{AllenJacob1985}, after we make the transformation 
by using the prescription given above.
To compare the results for arbitrary $\lam$, we 
consider the expression given in \cite{Youssef1}.
However the gauge parameter $\lam$ there is inverse of the
gauge parameter $\lam$ used by us. Keeping this 
under track, it is noticed that their expression 
exactly matches with our results.

\section{Summary and Conclusions}
\label{conc}

In this paper we computed the propagator for vector fields on
DeSitter background. We considered two cases of interest: massive 
and massless vector fields. For completeness we started with 
a discussion on bi-tensors in general\footnote{We used Synge's world function, the bi-scalar
$\sg(x, x^\prime)$, in our study of green's function. This should 
be contrasted with the all past studies done in the area,
where the bi-scalar $\mu(x, x^\prime)$ was used, which is 
actually proportional to square-root of the geodetic interval 
$\sg(x, x^\prime)$. We decided to work with $\sg(x, x^\prime)$
as it is always real (irrespective of the signature of metric), 
unlike $\mu(x, x^\prime)$ which can become complex in 
Lorentzian space-times. We have also not embedded deSitter space-time
in flat space-time by going to one higher dimension, thereby 
doing the analysis using a higher dimensional world function,
which has been done in \cite{Candelas1975}. In this sense we 
differ from \cite{Candelas1975}.}. 
This we covered in section \ref{Bi-tensor}\footnote{Although there is nothing
new in this section, but for completeness we decided to 
incorporate it in the manuscript for the ease of the readers.} 
After this we studied bi-tensors on a 
maximally symmetric space-times. Bi-tensors on a maximally 
symmetric background have been previously studied \cite{Katz, Peters}.
Here we made use of the important work done in \cite{Katz, Peters},
to write the arbitrary bi-tensors on maximally symmetric space-time
as a linear combination of bi-tensors 
constructed using $\sg_\mu$, $\sg_{\mu^\prime}$ and $g_\mnp$. 
We computed the expression for $\sg_\mn$, $\sg_\mnp$ and $g_{\mnp;\rho}$
on a maximally symmetric background. These are 
basic tensors that are needed in our analysis of green's function. 
More complicated bi-tensors can be constructed by making use of them.
Some of the more complicated ones that are used in the 
computation of green's function have been considered in the 
appendix \ref{useID}.

We then proceeded to compute the propagator of the vector fields in the massive 
and massless case respectively. During the computation we kept in mind the
following points, 
\begin{itemize}
\item At short distance the propagator on a DeSitter background 
should match the known result for the propagator on a flat space-time.
\item At large distance the propagator should be regular.
\end{itemize}
These considerations helped us in choosing the right vacuum 
on the deSitter background, which is the Bunch-Davis vacuum 
\cite{Bunch1978}, and the propagators are computed in this vacuum. 
This strategy of choosing right vacuum is well know and has been 
used by many in the past \cite{AllenJacob1985, Tsamis2006} (and references therein). 
Here we just follow some of those footsteps, and give a more unified 
treatment to both massive and massless vector fields. 

We use the path-integral language to 
write the equation satisfied by the green's functions (check eq. (\ref{eq:GW_iden_func}, 
\ref{eq:GRMV_curved}, and \ref{eq:eqmotion_U1})).
At the tree level and cases without interaction, this methodology
matches with the style of computing via equation of motion,
but deviations starts to occur when interactions are present
and when full exact propagator is computed.
In either case of massive and massless vectors we start by
analysing the correlation $\langle F_\mn F^{\mu^\prime\nu^\prime} \rangle$. 
This being a gauge invariant quantity, depends only on the 
transverse part of the green's function. 
We compute an additional quantity 
$\nb^\mu \langle F_\mn F^{\mu^\prime\nu^\prime} \rangle$ using the equation 
satisfied by the connected two-point correlation function. 

We decompose the full green's function $\langle A_\nu(x)  A_{\nu^\prime}(x^\prime) \rangle$
in to transverse and longitudinal parts. They are given by $G^T_\mnp$ and 
$\nb_\mu \nb_{\nu^\prime} G$ respectively, where the transverse part 
satisfies the condition $\nb^\mu G^T_\mnp=0$. It is shown that the transverse 
and longitudinal parts are obviously related to each other, due to the equation 
satisfied by the full green's function. In the massive case we get an extra 
constraint on the green's function $\nb^\nu \langle A_\nu(x)  A_{\nu^\prime}(x^\prime) \rangle =0$ 
(however this constraint doesn't imply that the longitudinal 
part of the green's function is zero, as is taken to be the case in
\cite{Tsamis2006}).
This is similar to Landau gauge in massless theories. 
We decide to solve for the transverse part of the green's function 
in both the massive and massless case. 
On a maximally symmetric background we write 
the transverse part of propagator as linear 
combination of $g_\mnp$ and $\sg_\mu\sg_{\nu^\prime}$,
with coefficients $\al$ and $\bt$ respectively being functions 
of $\sg$. Then we consider the 
quantity $\langle F_\mn F^{\mu^\prime\nu^\prime} \rangle$ on a maximally 
symmetric background, where it can be written as a linear combination 
of $g_{[\mu}{}^{[\mu^\prime}g_{\nu]}{}^{\nu\prime]}$
and $\sg_{[\mu}\sg^{[\mu^\prime}g_{\nu]}{}^{\nu\prime]}$, with the 
coefficients named $\ta$ and $\tau$ respectively. These coefficients 
are related to $\al$ and $\bt$ through the relations given in 
eq. (\ref{eq:albt_MV}). By using the equation satisfied 
by $\nb^\mu \langle F_\mn F^{\mu^\prime\nu^\prime} \rangle$, we find 
the differential equation satisfied by the function $\ta$, which when 
written after a change of variable from $\sg$ to $z$ (as defined in 
eq. (\ref{eq:zdef})), acquires a recognisable form of 
differential equation for the hypergeometric function. This is a second 
order differential equation and has two linearly independent solutions.
The right solution is picked by demanding that the solution should be singular 
at short distance and regular in the $z\to0$ limit. This picks one of the
linearly independent solution. The proportionality constant is fixed by
demanding that the strength of singularity of the solution in deSitter
should match the strength of singularity in flat space-time. 
Once $\ta(z)$ is obtained, it is used to find the function $\tau(z)$
using eq. (\ref{eq:tauZeq}). Finding these two functions determines 
the quantity $\langle F_\mn F^{\mu^\prime\nu^\prime} \rangle$ completely.
This is then used to determine $\al$ and $\bt$. As the transverse green's
function satisfies the transversality constraint $\nb^\mu G^T_\mnp=0$,
therefore this gives a relation between the function $\al$, $\bt$
and their derivatives. This is given in eq. (\ref{eq:albt_relt_MV}). 
From this $\bt$ and $\bt^\prime$ are eliminated using the first equation 
from (\ref{eq:albt_MV}), which ultimately gives an equation for the function
$\al$. This equation for both the massive (\ref{eq:aldiff_MV}) and massless
(\ref{eq:aleq_m0_GT}) case has the same structure, except the 
information stored in $\ta(\sg)$ on the RHS of these equation. 
These when written by making a change of variable from $\sg$ to $z$, 
is an in-homogenous second order differential equation of 
hypergeometric type. This has both particular and homogenous solution.
First we determine the particular solution, and then choose the 
homogenous solution. The coefficient in front of the homogenous 
solution is fixed by requiring that the leading singularity in both the
particular and homogenous solution should cancel each other.
Once the coefficient is fixed by using this requirement, it is noticed 
that the strength of next to leading singularity of the full solution 
for the transverse propagator matches the strength 
of singularity of the transverse propagator in flat space-time. 
Once the full solution for the function $\al(z)$ is found, it is used
to determine $\bt(z)$ by using eq. (\ref{eq:btsoln_MV}). This 
give us the knowledge of the full transverse propagator on 
deSitter background. In the massive case it is noticed that
the transverse propagator has a well-defined massless limit, 
in fact when $m^2 \to0$ the massive transverse propagator 
go to the massless transverse propagator (which is gauge 
independent). Here we wrote the expressions for the four
dimensions but this in true for arbitrary space-time dimensions. 

We then proceeded to determine the longitudinal part of the propagator.
The longitudinal part $\nb_\mu \nb_{\nu^\prime}G$ is first written 
as a linear combination of $g_\mnp$ and $\sg_\mu\sg_{\nu^\prime}$,
with the coefficients $\al_L$ and $\bt_L$ respectively, as in
eq. (\ref{eq:Gexp_longm0}). It is noticed that $\bt_L$ is 
related to $\al_L$ through the relation given in 
eq (\ref{eq:btL_inalL_m0}). So one needs to solve for $\al_L$
only, which can then be used to finally determine $G(z)$ and 
$\bt(z)$. In the case of massive vectors, the presence of 
constraint $\nb^\nu \langle A_\nu(x)  A_{\nu^\prime}(x^\prime) \rangle =0$, 
allows us to solve for the longitudinal part of the massive 
green's function. This generates an equation for the function 
$\al_L(\sg)$. This is given by eq. (\ref{eq:alL_MV_diff}). This should be
compared with the corresponding equation obtained for the massless 
case (\ref{eq:alLeq_m0}), where the RHS is gauge dependent. Putting 
$\lam=0$ in the eq. (\ref{eq:alLeq_m0}) shows that it exactly matches with the
massive case, indicating that the massive case is like considering 
Landau gauge. It is easy to solve the equation for $\al_L (z)$
for the massive case, as its a homogenous equation. While the 
corresponding equation for massless vector (\ref{eq:alLdiff_z_m0}) 
is non-homogenous and had both particular and homogenous solution
(and is dealt with in the same manner as is done for transverse part). 
In the massive case it is noticed that the solution for the longitudinal 
green's function has a $1/\g$ pole, meaning it diverges when mass
goes to zero. This is expected, as the same thing happens even in 
flat space-time also. As a result of this, the massless limit of the 
full massive propagator is ill defined, while the transverse part of the 
massive propagator is completely well behaved and goes to massless
transverse propagator in the massless limit, the same is not 
the case with the longitudinal part. One possible reason for this
sickness is the presence of zero modes, which should be 
subtracted from the propagator, to get a well defined 
green's function on the deSitter background. 

The interesting thing to note is the $z\to0$ limit of the propagators. 
The $z\to0$ limit of the massive transverse propagator matches 
with the $z\to0$ limit of the massless transverse propagator in the 
$m^2\to0$ limit. In either case of massive and massless 
propagators this correlation is negative and is proportional to the 
Ricci-scalar $R$, thereby vanishing when $R\to0$ (flat limit).  
In deSitter background, the non-vanishing of the two-point 
function in the antipodal limit indicates that there could be some edge 
states on the boundary whose contribution needs to be taken carefully. 
The $z\to0$ limit however is not the true infrared regime.
The euclidean deSitter due to its compact nature, allow 
the maximum separation between the two points to be 
antipodal. This is unlike the Lorentzian deSitter 
where the maximum separation between two points is infinite.

In the end we compare our results with the known expression for the
green's function for massive and massless vector fields. 
We note that we agree with the past literature on the known 
green's function for the massive and massless vector fields
on the deSitter background. I would like to mention that while 
are results may not be new, but our style of computation 
is different from what has been attempted in past, in the sense
that we compute the green's function by isolating the transverse 
part from the longitudinal part. This give us a benefit to the 
study the physical transverse part more carefully. It is seen that transverse 
part of both massive and massless theories are 
smoothly related to each other in the sense that transverse part 
of the massive vectors goes smoothly over to the ones of the 
massless theories in the massless limit. This comparison 
actually has been made at each level of our computation
and give us confidence that it is possible to give a unified 
treatment of the massive and massless theories. 
The longitudinal part however doesn't have this smooth 
transition. Isolating the longitudinal part helps us 
in investigating the famous infrared divergence problem
in the massive vector propagator when the mass goes to zero.
This is located in the longitudinal part of the propagator 
and is therefore unphysical. While this should not be a 
source of concern given that such problematic divergence 
is located in the longitudinal part, and is not expected to 
enter in any physical entity. But its presence does make 
us a bit uneasy and would have been nice if such 
issues were not present. 

The work can be generalised in several ways. 
The methodology described 
can be extended to take in to account interactions in a 
systematic way. It will be interesting to see whether in the 
presence of interaction there are formation of bound 
states or not. Finally it will be useful to translate this on 
a Lorentzian deSitter and investigate whether the 
propagator so obtained satisfies the various physical 
properties. Once the propagator of theory is obtained in Lorentzian 
space-time it will interesting to decipher the particle 
content from it. I plan to return to these issues in the 
subsequent paper.

\bigskip
\centerline{\bf Acknowledgements} 

GN would like to thank Ramesh Anishetty for several useful discussions. He would also like 
extend his thanks to Souradeep Majumder for his support during the course 
of this work. He would like to thank Prof. P. Spindel for pointing out 
\cite{Spindel1}. Finally I will like to extend my gratitude to IMSc
for hosting my visit and providing me wonderful hospitality, where a
part of the work was done. I am very grateful to some of my colleagues 
at IF for providing continuous support and encouragement during the 
course of this work.

\appendix

\section{Useful Identities}
\label{useID}

Here in this appendix we write some of the identities which are true 
on a maximally symmetric background and have been used 
while doing the computation. In this section we will be mostly using the 
expressions,
\bea
\label{eq:id1}
&&
\sg_{\mu\nu} = A(\sg) \biggl[g_\mn - \frac{1}{2\sg} \sg_\mu \sg_\nu
\biggr] + + \frac{1}{2\sg} \sg_\mu \sg_\nu \, \notag\\
&&
\sg_\mnp
= C(\sg) \biggl[g_\mnp + \frac{1}{2\sg} \sg_\mu \sg_{\nu^\prime}
\biggr] + \frac{1}{2\sg} \sg_\mu \sg_{\nu^\prime} \notag\\
&&
g_{\al\bt^\prime; \mu} = -\frac{A+C}{2\sg}[ g_{\mu\al} \sg_{\bt^\prime}
+ g_{\mu\bt^\prime} \sg_\al] \, .
\eea
Contracting $\mu$ and $\al$ in the last equation, we have
\beq
\label{eq:id2}
\nb^\mu g_\mnp = - \frac{A+C}{2\sg} (d-1) \sg_{\nu^\prime} \, .
\eeq
Acting with $\Box$ operator on $g_{\al\bt^\prime}$ we have,
\beq
\label{eq:id3}
\Box g_{\al\bt^\prime}
= \left(\frac{A+C}{2\sg} \right)^2 \left[
-2\sg g_{\al\bt^\prime} + (d-2) \sg_\mu \sg_{\nu^\prime}
\right]\, .
\eeq
By applying covariant derivative with respect to $x$ on $\sg_\mn$
and contracting it with one of the indices gives,
\beq
\label{eq:id4}
\sg_\mn{}^\nu = - \frac{(d-1) A(A-1)}{2\sg} \sg_\mu \, .
\eeq
Similarly applying covariant derivative with respect to 
$x$ on $\sg_\mnp$ and contracting the index with $\mu$
gives,
\beq
\label{eq:id5}
\nb^\mu \sg_\mnp
= \frac{(d-1)(A- C^2)}{2\sg} \sg_{\nu^\prime} \, .
\eeq
Now we consider the product of $\sg^{\mn}$ and $\sg_{\al\bt^\prime}$
and contract $\nu$ and $\al$. This gives after a bit of 
manipulation and using eq. (\ref{eq:id1}),
\beq
\label{eq:id6}
\sg_{\mu\rho} \sg^{\rho}{}_{\nu^\prime}
= A C g_\mnp
+ (1+AC) \frac{\sg_\mu \sg_{\nu^\prime}}{2\sg} \, .
\eeq
Now we can compute the expression of $\Box(\sg_\mu \sg_{\nu^\prime})$.
This can be computed as follows,
\beq 
\label{eq:id7}
\Box(\sg_\mu \sg_{\nu^\prime})
= \sg_{\nu^\prime} \Box \sg_\mu
+ \sg_\mu \Box \sg_{\nu^\prime}
+ 2 \sg_{\mu\rho} \sg^{\rho}_{\nu^\prime} \ , .
\eeq
Plugging the expression from the eq. (\ref{eq:id4},
\ref{eq:id5} and \ref{eq:id6}) in eq. (\ref{eq:id7}) we get,
\beq
\label{eq:id8}
\Box(\sg_\mu \sg_{\nu^\prime})
= 2 AC g_\mnp
+ \frac{1}{2\sg} \biggl[
2(d-1) A 
- (d-1) (A^2+C^2) +2+2AC
\biggr] \sg_\mu \sg_{\nu^\prime} \, .
\eeq
During the computation of Green function for the massless vector 
field, we will require the expressions for the following:
$\nb_{[\mu}\nb^{[\mu^\prime} g_{\nu]}{}^{\nu^\prime]}$, 
$\nb_{[\mu}\nb^{[\mu^\prime} (\sg_{\nu]}\sg^{\nu^\prime]})$.
They are given by following,
\bea
\label{eq:m0_iden1}
\nb_{[\mu}\nb^{[\mu^\prime} g_{\nu]}{}^{\nu^\prime]}
&=& \frac{C(A+C)}{2\sg} g_{[\mu}{}^{[\mu^\prime}g_{\nu]}{}^{\nu^\prime]}
+ \biggl[
\left(\frac{A+C}{2\sg} \right)^2
+ \frac{(C-1)(A+C)}{4\sg^2}
\notag \\
&&
+ \frac{A^\prime + C^\prime}{2\sg}
\biggr] \sg_{[\mu}\sg^{[\mu^\prime}g_{\nu]}{}^{\nu^\prime]} \, ,
\\
\label{eq:m0_iden2}
\nb_{[\mu}\nb^{[\mu^\prime} (\sg_{\nu]}\sg^{\nu^\prime]}
&=& -C^2 g_{[\mu}{}^{[\mu^\prime}g_{\nu]}{}^{\nu^\prime]}
- \frac{C(1+C)}{\sg} \sg_{[\mu}\sg^{[\mu^\prime}g_{\nu]}{}^{\nu^\prime]} \, .
\eea
Beside these we will also be needing the expression for 
$\nb^\mu (g_{[\mu}{}^{[\mu^\prime}g_{\nu]}{}^{\nu^\prime]})$
and $\nb^\mu (\sg_{[\mu}\sg^{[\mu^\prime}g_{\nu]}{}^{\nu^\prime]})$.
These can be computed easily and are given by following,
\bea
\label{eq:m0_iden3}
&&
\!\!\!\!\!\! \nb^\mu (g_{[\mu}{}^{[\mu^\prime}g_{\nu]}{}^{\nu^\prime]})
= -\frac{(d-2)(A+C)}{2\sg} \sg^{[\mu^\prime}
g_\nu{}^{\nu^\prime]} \, ,
\\
\label{eq:m0_iden4}
&&
\!\!\!\!\!\! \nb^\mu (\sg_{[\mu}\sg^{[\mu^\prime}g_{\nu]}{}^{\nu^\prime]})
= \frac{(d-2)A+2}{2} \sg^{[\mu^\prime}
g_\nu{}^{\nu^\prime]} \, .
\eea
Now we will like to compute $\nb_\mu\nb_\nu \left(g^\nu{}_{\nu^\prime}\right)$
and $\nb_\mu\nb_\nu \left(\sg^\nu \sg_{\nu^\prime} \right)$. They will 
occur during the computation of the green's function for the massless
vector fields. They are given by,
\bea
\label{eq:m0_iden5}
\nb_\mu\nb_\nu \left(g^\nu{}_{\nu^\prime}\right)
&=& - \frac{(d-1) C(A+C)}{2\sg} g_\mnp \, ,
\\
\label{eq:m0_iden6}
\nb_\mu\nb_\nu \left(\sg^\nu \sg_{\nu^\prime} \right)
&=& [(d-1)(2A-C^2+AC) 
+ 2(1+C)] \frac{\sg_\mu \sg_{\nu^\prime}}{2\sg}
\notag \\
&&
+ C[(d-1)A+2] g_\mnp \, .
\eea

\section{Massive Scalar in Flat Space-time}
\label{massscalar}

The path-integral for the massive scalar theory in 
euclidean space-time is given by,
\beq
\label{eq:LortoEuc}
Z_E = \int {\cal D} \phi_E \exp\biggl[
-\int \, {\rm d}^dx \biggl(
\frac{1}{2} \pt_\mu \phi_E \pt^\mu \phi_E
+ \frac{1}{2} m^2 \phi_E^2
\biggr)
\biggr] \, ,
\eeq
where the field $\phi_E(x)$ is the scalar field living in the 
euclidean flat space-time, with the positive signature. 
If a source is added to the euclidean path-integral, 
then integrating over the scalar field $\phi_E$ 
in flat space-time, we obtain an expression for the 
source dependent path-integral for the free scalar 
field in flat euclidean space-time (the same can be 
obtained for curved space by replacing the flat metric with 
curved one). This is given by,
\beq
\label{eq:ZJ_E0}
Z_E[J] = {\rm const.}\, \exp\biggl[
\frac{1}{2} \int \, {\rm d}^dx J(x) \D_F^{-1} J(x)
\biggr] \, ,
\eeq
where $\D_F = -\Box + m^2$. Taking derivatives of $Z[J]$ with 
respect to source twice, we obtain an expression for the 
two-point function for the scalar field. This is given by,
\beq
\label{eq:2pt_E}
G_E(x, x^\prime) = \langle \phi_E(x) \phi_E(x^\prime) \rangle
= \left. \frac{1}{Z} \frac{\de^2 \, Z}{\de J(x) \, \de J(x^\prime)} 
\right|_{J=0}
= \D_F^{-1} \de(x-x^\prime) \, ,
\eeq
Plugging the Fourier transform of the delta-function
give us the two-point function in Fourier space, having the 
following familiar form,
\beq
\label{eq:prop0_fourier}
G_E(x, x^\prime) = \int \frac{{\rm d}^dp_E}{\pd} 
\frac{e^{ip_E(x-x^\prime)}}{p_E^2 + m^2} \, ,
\eeq
where $p_E$ is the momentum in the euclidean space.
From now on for notational convenience we will be omitting 
the subscript $E$, as throughout this and the following 
appendix we will be dealing with euclidean Green's function. 

Now we make use of the inverse Laplace transform to re-write the 
reciprocal of $p^2+m^2$ as follows,
\beq
\label{eq:invlap_0}
\frac{1}{p^2+m^2}
= \int_0^{\infty} {\rm d} s \, e^{-s(p^2+m^2)} \, .
\eeq
Plugging this in the expression for propagator 
in eq. (\ref{eq:prop0_fourier}), and integrating 
over $p$, we obtain a neat expression for the 
propagator for the free massive scalar in configuration 
space with euclidean signature,
\beq
\label{eq:free0_prop}
G(x, x^\prime) = \frac{1}{(4\pi)^{d/2}} \int_0^\infty {\rm d}s \, 
\frac{1}{s^{d/2}} \exp \biggl[-\biggl(
s m^2 + \frac{\sg}{2s} \biggr)
\biggr] \, ,
\eeq
where $\sg = (x-x^\prime)^2/2$.
Now as we interested in the small mass limit, therefore we 
expand the integrand in powers of $m^2$. This allows us to 
integrate over $s$, and we get the following expression 
for the Green's function in euclidean space for the free 
massive scalar,
\beq
\label{eq:freemass_GE}
G(x, x^\prime) = \frac{1}{(4\pi)^{d/2}} 
\biggl(\frac{\sg}{2}\biggr)^{1-d/2}
\biggl[
\G \left(\frac{d}{2}-1\right)
- \frac{m^2 \sg}{2} \G \left(\frac{d}{2}-2\right)
+ \frac{m^4\sg^2}{8} \G \left(\frac{d}{2}-3\right)
+\cdots
\biggr] \, ,
\eeq
where $\G(x)$ is the Euler-Gamma function. 
From this we notice that in four dimensions, in the $\sg\to0$ limit
the leading term is singular, while other terms are finite. Also the 
leading term is independent of the mass of the scalar field. 
The massless limit is regular. 

\section{Massive Vector in Flat Space-time}
\label{massvec_flat}

Now we consider the case of massive vector field and compute the
propagator of the theory in flat euclidean space-time. The action
for this theory in curved euclidean space-time is given in 
eq. (\ref{eq:act_photon_c}) with the corresponding path-integral 
written in eq. (\ref{eq:MV_pathint}). The path-integral in 
eq. (\ref{eq:MV_pathint}) being gaussian in nature, can be 
performed easily thereby giving,
\beq
\label{eq:ZJ_E1}
Z[J] = {\rm const.} \, \exp\biggl[
\frac{1}{2} \int {\rm d}^dx J_\mu(x) \left(\D^{-1}_F\right)^\mn J_\nu(x)
\biggr] \, ,
\eeq
where $\left(\D_F\right)^\mn = (-\Box+m^2) \de^\mn + \pt^\mu \pt^\nu$. 
Taking two derivative with respect to the source $J_\mu$ at the point 
$J_\mu=0$ will give the two-point function, in other words the 
Green's function of the theory. This Green's function satisfies the 
following equation,
\beq
\label{eq:MVGF_E1}
\left[(-\Box+m^2) \de^\mn + \pt^\mu \pt^\nu \right]
\langle A_\nu A_{\nu^\prime} \rangle 
= \eta^\mu{}_{\nu^\prime} \de(x-x^\prime) \, ,
\eeq
where $\eta^\mu {}_{\nu^\prime}$ is flat space-time parallel 
displacement bi-vector, while the Green's function is given 
by $G_{\nu\nu^\prime} = \langle A_\nu A_{\nu^\prime} \rangle$. Using 
eq. (\ref{eq:MVGF_E1}) after doing some manipulation it 
is easy to obtain an expression for $\pt^\mu \langle F_\mn A_{\nu^\prime} \rangle$.
This is given by,
\beq
\label{eq:MV_FA_E1}
\pt^\mu \langle F_\mn A_{\nu^\prime} \rangle
= m^2 \langle A_\nu A_{\nu^\prime} \rangle - g_{\nu\nu^\prime} \de(x-x^\prime) \, .
\eeq
From this it is easily noticed that contracting the whole eq. (\ref{eq:MV_FA_E1})
with $\pt^\nu$, the LHS vanishes identically (due to anti-symmetry of 
$F_\mn$), while the RHS gives a constraint that need to be satisfied 
by the full Green's function. This constraint is given by,
\beq
\label{eq:MV_constr_flat}
\pt^\nu \langle A_\nu A_{\nu^\prime} \rangle
= \frac{1}{m^2} \eta_{\nu\nu^\prime} \pt^\nu \de(x-x^\prime) \, .
\eeq
It should be noticed that had we considered the case of 
non-coincident points then we would have got the 
constraint $\pt^\nu \langle A_\nu A_{\nu^\prime} \rangle=0$. 
It is crucial to retain the delta-function contribution in
eq. (\ref{eq:MV_constr_flat}) as it then gives the correct 
longitudinal part of the green's function. The green's function
however can be obtained via an alternative procedure. 
In configuration space, it is obtained by taking double 
derivative of the path-integral in eq. (\ref{eq:ZJ_E1}) with respect 
to the source $J_\mu$, where after the limit $J_\mu=0$ is taken in the end.
This gives the following form of the Green's function in the 
configuration space,
\beq
\label{eq:MV_E1}
G_\mnp (x, x^\prime) = g_{\nu\nu^\prime}
\left(\D_F^{-1}\right)_\mu{}^\nu \, \de(x - x^\prime) \, .
\eeq
Plugging the Fourier transform of the delta function yields an 
expression for the propagator in the momentum space to be,
\bea
\label{eq:MV_momE1}
G_{\nu\nu^\prime}(x, x^\prime)
= g_{\nu\nu^\prime} \int \frac{{\rm d}^dp}{\pd}
\biggl[
\frac{\de^\nu{}_\mu}{p^2 + m^2} 
+ \frac{p_\mu p^\nu}{m^2(p^2+m^2)} 
\biggr] e^{ip(x-x^\prime)} \, .
\eea
At this point it can be verified that this expression for the 
Green's function in momentum space satisfies the 
constraint stated in eq. (\ref{eq:MV_constr_flat}). 
This can be rewritten in a more compact form by making use
of the Green's function of the massive scalar
given in eq. (\ref{eq:prop0_fourier}). This is given by,
\beq
\label{eq:MV_E10rel}
G_\mnp(x, x^\prime)
= g_{\nu\nu^\prime} \biggl(
\de^\nu{}_\mu - \frac{\pt_\mu \pt^\nu}{m^2}
\biggr) G(x, x^\prime) \, .
\eeq
At this point we compute the quantity 
$\langle F_\mn F^{\mu^\prime\nu^\prime} \rangle$.
This quantity in flat space is given by,
\beq
\label{eq:FF_flat}
\langle F_\mn F^{\mu^\prime\nu^\prime} \rangle
= 4 \pt_{[\mu}\pt^{[\mu^\prime} \langle A_{\nu]} A^{\nu^\prime]}\rangle
= 4 g_{[\nu}{}^{[\nu^\prime} \pt_{\mu]} \pt^{\mu^\prime]} G(x, x^\prime) \, .
\eeq
From this we see that the $1/m^2$ term cancels from the expression 
due to anti-symmetry. Using the expression for $G(x, x^\prime)$ from 
eq. (\ref{eq:freemass_GE}), we can evaluate the leading terms
in the expansion of $\langle F_\mn F^{\mu^\prime\nu^\prime} \rangle$.
The first few terms of the series are given by,
\bea
\label{eq:FF_flat_series}
&&
\langle F_\mn F^{\mu^\prime\nu^\prime} \rangle
= \frac{1}{(4\pi)^{d/2}} \left(\frac{2}{\sg}\right)^{d/2+1} 
\biggl[
\G\left(\frac{d}{2}+1\right) 
\sg_{[\mu}\sg^{[\mu^\prime}g_{\nu]}{}^{\nu^\prime]}
+ \sg  \G\left(\frac{d}{2}\right) 
g_{[\mu}{}^{[\mu^\prime} g_{\nu]}{}^{\nu^\prime]}
\notag \\
&&
- \frac{m^2 \sg}{2}  \G\left(\frac{d}{2}\right) 
\sg_{[\mu}\sg^{[\mu^\prime}g_{\nu]}{}^{\nu^\prime]}
- \frac{m^2 \sg^2}{2} \G\left(\frac{d}{2}-1\right) 
g_{[\mu}{}^{[\mu^\prime} g_{\nu]}{}^{\nu^\prime]}
+ \cdots
\biggr] \, .
\eea
From this series expansion we notice that the leading term 
proportional to $g_{[\mu}{}^{[\mu^\prime} g_{\nu]}{}^{\nu^\prime]}$
is given by,
\beq
\label{eq:FF_flat_lead}
\langle F_\mn F^{\mu^\prime\nu^\prime} \rangle
\sim \frac{1}{(4\pi)^{d/2}}
\G\left(\frac{d}{2}\right) \left(\frac{2}{\sg}\right)^{d/2}
g_{[\mu}{}^{[\mu^\prime} g_{\nu]}{}^{\nu^\prime]} \, .
\eeq
This can be compared with the short distance limit of the function
$\ta(z)$ computed for the massive vector field on a DeSitter 
background in section \ref{massvec}.

From the expression for $G(x, x^\prime)$ in eq. (\ref{eq:freemass_GE}),
one can also work out the flat space-time propagator for the 
massive vector field using the expression stated in 
eq. (\ref{eq:MV_E10rel}). This is given by,
\beq
\label{eq:MV_E1expflat}
G_{\nu\nu^\prime}(x, x^\prime)
= \frac{1}{(4\pi)^{d/2}}
\left(\frac{2}{\sg}\right)^{d/2} \G\left(\frac{d}{2}+1\right) 
\biggl[
\frac{\sg_\mu \sg_{\nu^\prime}}{2m^2 \sg} 
+ \frac{g_\mnp}{d \, m^2}  
- \frac{\sg_\mu \sg_{\nu^\prime}}{2d} 
+ \frac{\sg g_\mnp}{d(d-2)} 
+\cdots
\biggr] \, .
\eeq
From this we notice that the leading singularity for $m^2 \to0$ 
and for small $\sg$ are the first two terms of the series expansion.
However this singularity which arises in the massless limit is 
longitudinal in nature. This can be verified by explicitly computing the 
longitudinal part of the green's function in flat space-time. This is 
given by,
\bea
\label{eq:GmnpLMV}
G_\mnp^L &=& g_\nu{}^{\nu^\prime} \frac{\pt^\nu \pt_\mu}{m^2} 
\int \frac{{\rm d}^d p}{\pd} \frac{e^{\imath p(x-x^\prime)}}{p^2} 
= g_\nu{}^{\nu^\prime} \frac{\pt^\nu \pt_\mu}{m^2} G(\sg) 
\notag \\
&=& -\frac{1}{m^2} \frac{1}{(4\pi)^{d/2}} 
\left(\frac{2}{\sg} \right)^{d/2+1} \G\left(\frac{d}{2} +1 \right)
\biggl[
\frac{1}{4} \sg_\mu \sg_{\nu^\prime}
+ \frac{\sg}{2 d}  g_\mnp
\biggr] \, ,
\eea
where $G(\sg)$ is the euclidean green's function for the massless
scalar in flat space-time. Its expression is given by following,
\beq
\label{eq:GF_m0flat_scalar}
G(\sg) = \int \frac{{\rm d}^dp}{\pd}
\frac{e^{ip(x-x^\prime)}}{p^2} 
= \frac{1}{(4\pi)^{d/2}} 
\biggl(\frac{2}{\sg}\biggr)^{\frac{d}{2}-1} 
\G\left(\frac{d}{2} -1 \right) \, .
\eeq
Removing the longitudinal part from the full 
green's function give us the transverse part of the 
propagator in flat space-time. This is given by,
\beq
\label{eq:GmnpTMV}
G^T_\mnp = \frac{1}{(4\pi)^{d/2}} \biggl[
-\frac{\sg_\mu \sg_{\nu^\prime}}{4} 
\G\left(\frac{d}{2} \right) \left(\frac{\sg}{2} \right)^{-d/2}
+ \frac{g_\mnp }{2} \G\left(\frac{d}{2} -1\right) 
\left(\frac{\sg}{2} \right)^{1-d/2}
+\cdots
\biggr] \, .
\eeq
It is noticed that the transverse part of the green's function has 
a well defined massless limit. However the longitudinal part 
suffers from a singularity in the limit $m^2\to0$. But it should
also be stressed that the longitudinal part here is non-propagating.
The transverse and longitudinal part of the Green's function 
in flat space-time will be used as boundary condition 
to correctly determine the transverse and longitudinal part 
of the Green's function on the deSitter background.

\section{Massless Vector in Flat Space-time}
\label{masslessvecflat}

Now we consider the case of massless vector fields. Again we will 
work in the euclidean signature. 
We start by considering the path integral for the 
massless vector field which has been gauge fixed. In the 
euclidean signature this is given by,
\beq
\label{eq:pathint_m0_E}
Z[J] = \int {\cal D} A_\mu
\exp\biggl[
- \int {\rm d}^dx \biggl(\frac{1}{4} F_\mn F^\mn 
+ \frac{1}{2\lam} (\pt_\mu A^\mu)^2 \biggr)
- \int {\rm d}^dx J_\mu A^\mu
\biggr] \, ,
\eeq
where $J_\mu$ is the source. One can obtain various correlation 
function of the field $A_\mu$ by taking successive derivatives of the
path-integral with respect to the source $J$. In particular two derivatives
will give the two-point function of the theory. Being a free theory 
this will be the tree-level green's function of the theory. The path-integral
being quadratic in the gauge field allows one to perform the gaussian type 
integral thereby giving an expression for the path-integral explicitly 
in terms of source and operators. This is given by,
\beq
\label{eq:pathint_m0_exp}
Z[J] = {\rm const.} \exp
\biggl[
\frac{1}{2} \int {\rm d}^dx J_\mu 
\left(\D_F^{-1} \right)^\mn J_\nu
\biggr] \, .
\eeq
where $\left(\D_F\right)_\mn = -\de_\mn \Box
+ (\lam-1)/\lam \pt_\mu \pt_\nu$. Then the 
green's function for the massless vector is given by,
\beq
\label{eq:green_flat_m0}
G_\mnp (x, x^\prime)
= g_{\nu\nu^\prime} \left(\D_F^{-1} \right)_\mu{}^\nu \de(x-x^\prime) \, .
\eeq
At this point one can plug the Fourier transform 
of the $\de$-function in the expression and obtain the 
green's function of the massless vector field in the 
momentum space to be,
\beq
\label{eq:GF_m0flat_mom}
G_\mnp (x, x^\prime)
= g_{\nu\nu^\prime} \int \frac{{\rm d}^dp}{\pd}
\biggl[
\frac{1}{p^2} \de_\mu^\nu
+ (\lam-1)\frac{p_\mu p^\nu}{p^4} 
\biggr] e^{ip(x-x^\prime)} \, .
\eeq
This can be rewritten by pulling out the derivative acting on $x$,
thereby giving the expression for the green function for massless 
vector in terms of the green's function for massless scalar. 
\beq
\label{eq:GF_m0_pullder}
G_\mnp (x, x^\prime)
= g_{\nu\nu^\prime} \biggl[
\de_\mu^\nu +(\lam-1)
\frac{\pt_\mu \pt^\nu}{\pt^2}
\biggr] \int \frac{{\rm d}^dp}{\pd}
\frac{e^{ip(x-x^\prime)}}{p^2} \, .
\eeq
Now from this expression one can compute the expectation value
$\langle F_\mn F^{\mu^\prime\nu^\prime} \rangle$. This is given by,
\beq
\label{eq:FF_flat_m0}
\langle F_\mn F^{\mu^\prime\nu^\prime} \rangle
= 4 g_{[\nu}{}^{[\nu^\prime} \pt_{\mu]}\pt^{\mu^\prime]}
G(\sg) \, ,
\eeq
where $G(\sg)$ is given in eq. (\ref{eq:GF_m0flat_scalar}), using which 
in eq. (\ref{eq:FF_flat_m0}) we obtain,
\beq
\label{eq:FF_flatm01}
\langle F_\mn F^{\mu^\prime\nu^\prime} \rangle
= \frac{1}{(4\pi)^{d/2}} \G\left(\frac{d}{2} +1 \right) 
\biggl(\frac{2}{\sg}\biggr)^{\frac{d}{2}+1} 
\biggl[
\sg_{[\mu}\sg^{[\mu^\prime} g_{\nu]}{}^{\nu^\prime]}
+ \frac{2}{d \, \sg} g_{[\mu}{}^{[\mu^\prime}g_{\nu]}{}^{\nu^\prime]} 
\biggr] \, .
\eeq
To compute the quantity $\langle A_\mu A_{\nu^\prime} \rangle$ instead 
of using the expression stated in the eq. (\ref{eq:GF_m0_pullder})
(as it involves inverse of laplacian) we adopt a different strategy.
We rewrite the eq. (\ref{eq:GF_m0_pullder}) in a slightly different
form as,
\beq
\label{eq:flatprop_MSV}
G_\mnp (x, x^\prime)
= -g_{\nu\nu^\prime} \biggl[
\de_\mu^\nu \pt^2 + (\lam-1)
\pt_\mu \pt^\nu
\biggr] \int \frac{{\rm d}^dp}{\pd}
\frac{e^{ip(x-x^\prime)}}{p^4} \, .
\eeq
Now the quantity under the integral can rewritten using the 
inverse Laplace transform as,
\beq
\label{eq:InvLap_AA_MSV}
G_\mnp (x, x^\prime)
= g_{\nu\nu^\prime} \biggl[
\de_\mu^\nu \pt^2 + (\lam-1)
\pt_\mu \pt^\nu
\biggr]  
\int \frac{{\rm d}^dp}{\pd}
\int_0^{\infty} {\rm d} s \, s e^{-s p^2+ip(x-x^\prime)}\, ,
\eeq
At this point one can perform the integral over the momentum by
completing the square, and then perform the integration of the 
variable $s$ by making use of the definition of the Euler-Gamma 
functions. Then after a bit of manipulation we will get the following expression,
\beq
\label{eq:AA_MSV_flat}
G_\mnp (x, x^\prime)
= g_{\nu\nu^\prime} \frac{1}{(4\pi)^{d/2}} 
\G\left(\frac{d}{2}-2\right)
\biggl[
\de_\mu^\nu \pt^2 + (\lam-1)
\pt_\mu \pt^\nu
\biggr] \left(\frac{\sg}{2} \right)^{2-d/2} \, .
\eeq
This can be written in transverse and longitudinal parts
by rewriting the derivative part in the following way,
\beq
\label{eq:AA_TVdecomp}
G_\mnp (x, x^\prime)
= g_{\nu\nu^\prime} \frac{1}{(4\pi)^{d/2}} 
\G\left(\frac{d}{2}-2\right)
\biggl[
(\de_\mu^\nu \pt^2 - \pt_\mu \pt^\nu) + \lam
\pt_\mu \pt^\nu
\biggr] \left(\frac{\sg}{2} \right)^{2-d/2} \, .
\eeq
Now one can perform the derivative operation easily 
without running in to problems involving inverse of 
flat space-time laplacian operator.
This gives the propagator in flat space-time for the photon field to be,
\beq
\label{eq:AA_prop_flat}
G_\mnp (x, x^\prime)
= \frac{1}{(4\pi)^{d/2}} \left(\frac{\sg}{2} \right)^{1-d/2}
\G\left(\frac{d}{2}\right)
\biggl[
\biggl\{
\frac{g_\mnp}{d-2} - \frac{\sg_{\mu}\sg_{\nu^\prime}}{2\sg}
\biggr\}
+ \lam
\biggl\{
\frac{g_\mnp}{d-2} + \frac{\sg_{\mu}\sg_{\nu^\prime}}{2\sg}
\biggr\}
\biggr] \, .
\eeq
Here the term independent of $\lam$ is the transverse 
part of propagator while the term proportional to $\lam$
is the longitudinal part.
In the Landau gauge ($\lam=0$) the longitudinal part disappears 
giving only the contribution from the transverse physical part.
It should be noticed that the transverse part here for the massless 
propagator matches with the transverse part of the massive propagator
in flat space time given in eq. (\ref{eq:GmnpTMV}).


\end{document}